\documentclass[nohyper]{JHEP3}

\usepackage{amsmath,amssymb}
\usepackage{graphicx}
\usepackage{cite}

\newcommand{\lsim}
{\raise0.3ex\hbox{$\;<$\kern-0.75em\raise-1.1ex\hbox{$\sim\;$}}}
\newcommand{\gsim}
{\raise0.3ex\hbox{$\;>$\kern-0.75em\raise-1.1ex\hbox{$\sim\;$}}}

\newcommand{\be}{\begin{equation}}
\newcommand{\ee}{\end{equation}}
\newcommand{\bea}{\begin{eqnarray}}
\newcommand{\eea}{\end{eqnarray}}

\renewcommand{\thefigure}{\Roman{figure}}
\newcommand{\eq}[1]{Eq.~(\ref{#1})}

\newcommand{\ADi}{\left|A_D^i\right|^2}
\newcommand{\ANLH}{\left| {A} ({N_i} \to {L} {H} )\right|^2} 
\newcommand{\ALHN}{\left| {A} (L H \to {N_i})\right|^2} 
\newcommand{\ANLHb}{\left| {A} ({N_i} \to \bar{L} \bar{H} )\right|^2} 
\newcommand{\ALHNb}{\left| {A} (\bar{L} \bar{H} \to {N_i})\right|^2} 
\newcommand{\ANLHal}{\left| {A} ({N_i} \to {L_\alpha} {H} )\right|^2} 
\newcommand{\ALHNal}{\left| {A} (L_\alpha H \to {N_i})\right|^2} 
\newcommand{\ANLHbal}{\left| {A} ({N_i} \to \bar{L}_\alpha \bar{H} )\right|^2} 
\newcommand{\ALHNbal}{\left| {A} (\bar{L}_\alpha \bar{H} \to {N_i})\right|^2} 
\newcommand{\ANLHbbet}{\left| {A} ({N_i} \to \bar{L}_\beta \bar{H}) \right|^2} 
\newcommand{\e}{\bar{E}} 
\newcommand{\sL}{\widetilde{L}} 

\newcommand{\epi}{\epsilon_{i}} 
\newcommand{\epial}{\epsilon_{i}^{\alpha}} 
\newcommand{\efi}{\epsilon_{f_i}}
\newcommand{\esi}{\epsilon_{s_i}} 

\newcommand{\Nif}{|\hat{A} ( \widetilde{N}_i \to L h  )|^2}

\newcommand{\Nis}{|\hat{A}( \widetilde{N}_i
\to \widetilde{L} H )|^2} 

\newcommand{\Nifb}{
| \hat{A} (\widetilde{N}_i \to \bar{L}
\bar{h} )|^2} 

\newcommand{\Nisb}{|\hat{A}(
\widetilde{N}_i \to \widetilde{L}^{\dag}H^{\dag})|^2}

\newcommand{\ffNi}{
| \hat{A} (L h \to\widetilde{N}_i)|^2} 

\newcommand{\fbNi}{| \hat{A}
(\bar{L} \bar{h} \to \widetilde{N}_i )|^2}

\newcommand{\sNi}{
|\hat{A}( \widetilde{L} H \to \widetilde{N}_i )|^2}

\newcommand{\sbNi}{
| \hat{A}( \widetilde{L}^{\dag} H^{\dag} \to
\widetilde{N}_i)|^2} 

\newcommand{\Af}{|A^{f}_i|^2}

\newcommand{\As}{|A^{s}_i|^2}
\newcommand{\esm}{ \epsilon_{s-} } 
\newcommand{\esp}{\epsilon_{s+}}
\newcommand{\efm}{ \epsilon_{f-} } 
\newcommand{\efp}{\epsilon_{f+}}

\newcommand{\fL}{f_L} 
\newcommand{\fLb}{f_{\bar{L}}} 
\newcommand{\fLt}{f_{\widetilde{L}}} 
\newcommand{\fLtb}{f_{\widetilde{L}^\dagger}}
\newcommand{\flep}{f_{\cal{L}}}
\newcommand{\fslep}{f_{\widetilde{\cal{L}}}}
\newcommand{\fLeq}{f_L^{eq}} 
\newcommand{\fheq}{f_h^{eq}}
\newcommand{\fLteq}{f_{\widetilde{L}}^{eq}}
\newcommand{\fHeq}{f_H^{eq}}
\newcommand{\fNieq}{f_{N_i}^{eq}}
\newcommand{\fsNieq}{f_{\widetilde{N}_i}^{eq}}
\newcommand{\fsNi}{f_{\widetilde{N}_i}}
\newcommand{\fsNeq}{f_{\widetilde{N}}^{eq}}
\newcommand{\fsN}{f_{\widetilde{N}}}

\newcommand{\YLf}{Y_{\cal{L}}}
\newcommand{\YLs}{Y_{\widetilde{\cal{L}}}}
\newcommand{\YLT}{Y_{\cal{L}_T}}

\newcommand{\pN}{{p}_{N_i}}
\newcommand{\pL}{{p}_L}
\newcommand{\pH}{{p}_H}

\newcommand{\eps}{\epsilon}

\newcommand{\bolN}{\frac{\partial f_{N_i}}{\partial t}- p_{N_i} H 
\frac{\partial f_{N_i}}{\partial {p}_{N_i}}}
\newcommand{\bolsNi} {\frac{\partial f_{\widetilde{N}_i}}{\partial t}- 
{p}_{\widetilde{N}_i} H 
\frac{\partial f_{\widetilde{N}_i}}{\partial {p}_{\widetilde{N}_i}}}
\newcommand{\bollep}{\frac{\partial f_{\cal{L}}}{\partial t}- {p}_L H 
\frac{\partial f_{\cal{L}}}{\partial {p}_L}}
\newcommand{\bolslep}{\frac{\partial f_{\widetilde{\cal{L}}}}{\partial t}- 
{p}_L H \frac{\partial f_{\widetilde{\cal{L}}}}{\partial {p}_L}}
\newcommand{\bollepal}{\frac{\partial f_{\cal{L}_\alpha}}{\partial t}- {p}_L H 
\frac{\partial f_{\cal{L}_\alpha}}{\partial {p}_L}}

\title{On the full Boltzmann equations for Leptogenesis}

\author{J. Garayoa\footnote{garayoa@ific.uv.es}, 
S. Pastor\footnote{pastor@ific.uv.es},
T. Pinto, N. Rius\footnote{nuria@ific.uv.es}\,
and O. Vives\footnote{vives@ific.uv.es}  
\\
Depto.\ de F\'{\i}sica Te\'orica,
and IFIC, Universidad de
Valencia-CSIC \\ 
Edificio de Institutos de Paterna, Apt. 22085, 46071 Valencia,
Spain}

\abstract{We consider the full Boltzmann equations for standard and 
soft leptogenesis, 
instead of the usual integrated Boltzmann equations which assume kinetic
equilibrium for all species. Decays and inverse decays may be 
inefficient for thermalising the heavy-(s)neutrino distribution 
function, leading to significant deviations from kinetic equilibrium. 
We analyse the impact of using the full kinetic equations in the case of  
a previously generated lepton asymmetry, 
and find that the washout of this initial asymmetry due to the interactions 
of the right-handed neutrino 
is larger than when calculated via the integrated equations. 
We also solve the full Boltzmann equations for soft leptogenesis,  
where the lepton asymmetry induced by the soft SUSY-breaking terms in 
sneutrino decays is a purely thermal effect, since at $T=0$ the asymmetry 
in leptons cancels the one in sleptons. In this case, we obtain that 
in the weak washout regime ($K \lesssim 1$) the final lepton 
asymmetry can change up to a factor four with respect to
previous estimates.}  
\preprint{%
  IFIC/09-13 \\
  FTUV-09-0419}

\begin{document}

\section{Introduction}

The discovery of neutrino oscillations makes leptogenesis a very
attractive solution to the baryon asymmetry problem \cite{fy}. It is 
usually assumed that the tiny neutrino masses are generated via the
(type I) seesaw mechanism \cite{ss} and thus the new singlet neutral
leptons with heavy (lepton number violating) Majorana masses can
produce dynamically a lepton asymmetry through out of equilibrium
decays. Eventually, this lepton asymmetry is partially converted into a
baryon asymmetry due to fast $B-L$ violating sphaleron processes.

Most studies of leptogenesis use the integrated Boltzmann equations 
to follow the evolution of the heavy particle number density and 
the lepton asymmetry. This approach assumes Maxwell--Boltzmann statistics, 
as well as kinetic equilibrium for all particles, including  the heavy 
species.
This assumption is normally justified in freeze-out calculations, where
elastic scattering is assumed to be much faster than inelastic reactions.
However, in the present context, kinetic equilibrium for the heavy species 
would have to be maintained basically 
by the decays and inverse decays alone, and
it is not obvious that the integrated Boltzmann equation is always a 
good approximation. In general, $1 \leftrightarrow 2$ processes are 
more inefficient for thermalization compared to $2 \leftrightarrow 2$ 
processes, and in some parameter ranges there can be large deviations 
from kinetic equilibrium. 

In \cite{hann}, the impact of this difference
on the lepton asymmetry produced during leptogenesis was studied, 
and it was found that in the strong washout regime the final 
asymmetry is changed by $15-30\%$ when the full Boltzmann equations 
are used. 
In this work we extend the study to two different scenarios
not considered previously, in which the effects can be sizeable:

(i) Preexisting lepton asymmetry.
A lepton asymmetry that has been previously generated, for instance  
by the next-to-lightest right 
handed neutrino, $N_2$, tends to be washed out by 
the interactions of the lightest one, $N_1$. 
If $M_{N_1} \ll 10^9$ GeV, the lepton asymmetry generated in its decay can be
neglected and given that the $N_1$ are thermally produced at temperatures close
to $M_1$, 
this washout is exponential and therefore a change 
of order $20\%$ may be important. We have used the full Boltzmann  
equations to calculate the evolution of the lepton asymmetry, 
created during $N_2$ decay, at lower temperatures,  
$T \sim M_1 \ll M_2$.
We have also generalize them to the flavoured leptogenesis case. 

(ii) Soft leptogenesis.  
Since right-handed neutrino masses and therefore leptogenesis are usually
associated to a very high energy scale, 
a supersymmetric scenario is desirable in order to stabilize the hierarchy  
between the leptogenesis scale and the electroweak one.
It has been shown \cite{soft1a,soft1b,soft2} that supersymmetry-breaking 
terms can play an important role in the generation of a lepton asymmetry 
in sneutrino decays:  
they remove the mass degeneracy between the two real sneutrino states of  
a single neutrino generation, and also provide new sources of lepton number 
and CP violation.  
As a consequence, the mixing between the two sneutrino
states generates a CP asymmetry in the decay, which can be sizable  
because of the resonant effect \cite{resonant} of the two 
nearly-degenerate states. 
This scenario has been termed ``soft leptogenesis'', since the soft terms
and not flavour physics provide the necessary mass splitting and 
CP-violating phase.  It has also been studied in the minimal 
supersymmetric triplet seesaw model \cite{anna,cs} and in the 
inverse seesaw scenario \cite{ggr}.  
An important difference with respect to
the standard leptogenesis mechanism, 
is that
the lepton asymmetry produced through soft leptogenesis 
is a pure thermal effect, because at $T=0$ the asymmetry in leptons  
cancels the one in sleptons. Only at finite temperature the difference 
between the fermion 
and boson statistics leads to a non-vanishing lepton and CP asymmetry,
so in this case the use of the full Boltzmann equations is  
mandatory. Several approximations have been used in the literature
\cite{soft2,cs}, 
and we will compare our exact results with them.

This paper is organized as follows. In section 2, we derive the full  
Boltzmann equations for standard leptogenesis and we study the washout 
of a previously generated lepton asymmetry.
Section 3 is devoted to soft leptogenesis.
In section 4, we present our conclusions, and more technical details
concerning the full Boltzmann equations are described in the appendices.

\section{Standard leptogenesis}
\label{lepto} 

In this section, we review the full Boltzmann equations
relevant for leptogenesis, in a simplified model which  
includes only the heavy right-handed neutrino ($N_i$)
decays, inverse decays, and  
resonant scattering. The off-shell $2 \leftrightarrow 2$ scattering
processes mediated by $N_i$ have only small effects for $T<10^{12}$ GeV and 
can be neglected in first approximation \cite{flavour,pedestrians}.
We then investigate how the use of the full evolution equations affects 
the final lepton number asymmetry. 
We do not include thermal corrections \cite{thermal} 
and we consider only initial zero abundance of the heavy neutrinos.

The CP asymmetry in the decay of the  
right-handed neutrino $N_i$ is:  
\begin{eqnarray} 
\epi & = & \frac
{\left|A(N_i  \to L H) \right|^2-\left|A(N_i  \to \bar{L} \bar{H})\right|^2}
{\left|A(N_i  \to L H) \right|^2+\left|A(N_i  \to \bar{L} \bar{H})\right|^2} 
= \frac
{\left|A(N_i  \to L H) \right|^2-\left|A(N_i  \to \bar{L} \bar{H})\right|^2}
{\ADi} \ ,
\end{eqnarray}
where we implicitly sum over all flavours, since, at this point, we work in 
the single flavour approximation.    
Its decay width is
\be  
\label{gammai} 
\Gamma_i = \frac{\ADi}{16 \pi M_i} = \frac {M_i}{8 \pi} 
\sum_\alpha |Y_{\alpha i}|^2 \; ,
\ee  
where $Y_{\alpha i}$ are the Yukawa couplings of the heavy neutrinos.

We will denote $f_a$ the phase-space density of a particle species $a$,  
so its number density is given by 
\be 
n_a = g_a\int \frac{d^3 p}{(2\pi)^3} f_a(\bar{p}) \,  ,
\ee 
where $g_a$ is the number of internal degrees of freedom. 
In order to eliminate the dependence in the expansion of the Universe,  
as usual, we write the equations in terms of the abundances,  
$Y_a = n_a/s$, being $s$ the entropy density,
\be
\label{entropy} 
s= \frac {2 \pi^2}{45} g_* T^3 
\ee 
with $g_*$ the number of relativistic degrees of freedom at temperature 
$T$, so that $g_*=106.75$ in the Standard Model (SM) and $g_*=228.75$ in 
the Minimal Supersymmetric Standard Model (MSSM), which  
we consider in Sec.~\ref{slepto}.

We will study the time evolution of the right-handed neutrino distribution  
$f_{N_i}$, and the lepton asymmetry distribution  
$f_{\cal{L}}= f_L -f_{\bar{L}}$.
Due to the fast gauge interactions, to a good approximation 
the Higgs field and the leptons are in kinetic equilibrium.
Moreover, the Higgs number asymmetry is not conserved due to the large 
top Yukawa coupling, so we neglect the Higgs chemical 
potential\footnote{The effect of keeping the 
Higgs number asymmetry  
has been studied in \cite{spectators}, and could lead to 
a reduction of the final baryon asymmetry of ${\cal O}(1)$.}.
Then, we consider the following  distributions:
\begin{eqnarray} 
\label{distrH}
f_H^{eq} & =& (e^{E_H/T}- 1)^{-1} \, ,\\ 
\label{distrL}
f_L & =& (e^{(E_L-\mu)/T } + 1)^{-1} \, , \qquad  
f_{\bar{L}}  = (e^{(E_{\bar{L}}+\mu)/T} + 1)^{-1} \, ,
\end{eqnarray} 
where we have introduced a chemical potential for the leptons, $\mu$. 
Following \cite{hann}, we use that $\mu/T \ll 1$ and we approximate:
\be  
\label{flep}
f_{\cal{L}} =  \frac {2 \, e^{E_L/T }}{(e^{E_L/T } + 1)^2}\frac{\mu}{T} 
+ {\cal O}((\frac {\mu}{T})^3) \ ,
\ee  
\be 
\label{fl2} 
f_L  +f_{\bar{L}} \simeq 2 \fLeq + {\cal O}((\mu/ T)^2)  \, .
\ee 
where , 
\be
\fLeq   = (e^{E_L/T } + 1)^{-1}  \, .\\ 
\ee
Then,
the lepton asymmetry $Y_{\cal{L}}= n_{\cal{L}}/s$ is given by  
\be
\label{mu} 
Y_{\cal{L}} = \mu  \, \frac {T^2}{3s} + 
{\cal O}((\frac{\mu}{T})^3) \, .  
\ee 
Note that in this section we have neglected the thermal masses 
$m_H(T)$, $m_L(T)$,   
therefore  $E_{H,L}=|\bar{p}_{H,L}| \equiv p_{H,L}$.
A thorough study of thermal leptogenesis can be found in  
\cite{thermal}, where it has been shown that 
when thermal masses are taken into account,  
at sufficiently high temperature, the Higgs becomes heavier than 
$N_i$ and the decay $N_i \rightarrow H L$ is kinematically  
forbidden. At higher temperatures, the Higgs becomes heavy 
enough to allow the decay $H \rightarrow N_i L$, where also  
a CP asymmetry $\epsilon_H$ is produced. However this asymmetry 
turns out to have a negligible effect on the final results 
for leptogenesis, so we do not consider it here.

Using the formalism of appendix \ref{appA}, we can write  
the full Boltzmann equations for the heavy neutrino and the lepton asymmetry.
We assume that the later  is small, so  
we work to first order in $\epsilon$ and $Y_{\cal{L}}$. 
It is convenient to  
use the dimensionless variables $z_i=M_i/T$, $\e_i=E_{N_i}/T$ and 
$y_a= p_a/T$ ($a=N_i,H,L$) to eliminate the dependence on 
the expansion rate of the Universe, so we obtain
\footnote{Our equation for the evolution of $\flep$ is slightly different  
from the one in \cite{hann}, due to the fact that they add the term 
$2 \epsilon_i (1-f_{N_i})\fHeq  2\fLeq$ (in our notation) because of  
the resonant part of the $L H \leftrightarrow \bar{L} \bar{H}$ scattering.
However, this resonant contribution involves $\fNieq$ instead  
of $f_{N_i}$, as we show in the appendix, so we have an  
extra term $4 \epsilon_i \fHeq \fLeq (f_{N_i}-\fNieq)$. We have 
checked that this difference is numerically very small and does not  
change the main results.}: 
\bea  
 \label{ec_N_adim}  
\frac{\partial f_{N_i}}{\partial z_i}& =& \frac{K_i z_i^2}{y_i \e_i} 
\int_{\frac{\e_i - y_i}{2}}^{\frac{\e_i + y_i}{2}}   
dy_H \left[\fHeq \fLeq (1-f_{N_i}) - f_{N_i} (1-\fLeq)(1+\fHeq)  \right] \, , 
\\ 
\label{ec_L_adim} 
\frac{\partial \flep}{\partial z_i} & =  & \frac{K_i z_i^2}{y_L^2} 
\int_{y_L + \frac{z_i^2}{4 y_L}}^{\infty} d\e_i  
 \left\{ \epi \, (f_{N_i} - \fNieq)   
\left[(1-\fLeq)(1+\fHeq) - \fHeq \fLeq \right] 
 \right. - \nonumber \\   
 &  & \left. - \frac 1 2  f_{\cal{L}} (\fHeq+f_{N_i}) \right\} \ , 
\eea  
where we have defined the decay  parameter 
$K_i  \equiv \Gamma_i/H(T=M_i)$ , which controls  
whether or not $N_i$ decays are in equilibrium.

Equations (\ref{ec_N_adim}) and (\ref{ec_L_adim}) can be integrated
numerically for given values of $K_i$ and $\eps_i$. Note that within our 
first order calculation, in the evolution equation for the heavy
neutrino distribution \eq{ec_N_adim}  only the {\em 
   equilibrium} Higgs and lepton distributions appear, so it can be solved 
independently of the lepton asymmetry.  Using the approximate
expression  \eq{flep} for $\flep$,  
the integration of 
 \eq{ec_L_adim} over the dimensionless lepton momentum, $y_L$, 
leads to the following evolution equation for the chemical potential,
\bea
 \label{ec_mu} 
\frac 1 T \frac{d\mu}{dz_i} &= &\frac{3 K_i z_i^2}{\pi^2} 
 \int_{z_i}^{\infty} d\e_i 
\int_{\frac{\e_i - y_i}{2}}^{\frac{\e_i+y_i}{2}} dy_L
  \left\{   - \frac{\mu}{T} \frac{e^{y_L}}{(1+e^{y_L})^2}  (\fHeq+f_{N_i}) + 
 \right. \nonumber \\  
  &  &   \left. + 
 \epi \, (f_{N_i} - \fNieq)   
 \left[(1-\fLeq)(1+\fHeq) - \fHeq \fLeq \right] \right\} \ , 
\eea 
 Recall that $\mu$ is related to the lepton asymmetry $\YLf$ by \eq{mu}.

 It is straightforward to verify that if Maxwell--Boltzmann statistics and 
kinetic equilibrium for all species are assumed,  
so that the phase space distributions are
\be 
f_{N_i} = \frac{Y_{N_i}}{Y_{N_i}^{eq}} \, e^{-\e_i} \; ,  \qquad  
\fHeq = e^{-\e_H} \; , \qquad 
f_{L,\bar{L}} = e^{-\e_L \pm \mu/T} \; , 
\ee 
the above equations can be easily integrated and one 
recovers the usual integrated Boltzmann equations for leptogenesis: 
\bea \label{int_BESN} 
\frac{d Y_{N_i}}{dz_i} &=& - K_i z_i (Y_{N_i}-Y_{N_i}^{eq})  
\frac{{\cal K}_1(z_i)}{{\cal K}_2(z_i)} \, ,\\ 
\label{int_BESL} 
\frac{d \YLf}{dz_i} &=& \epi \, K_i z_i (Y_{N_i}-Y_{N_i}^{eq}) 
\frac{{\cal K}_1(z_i)}{{\cal K}_2(z_i)}  
- \frac{z_i^3}{4} K_i {\cal K}_1(z_i) \YLf \, ,
\eea  
where ${\cal K}_1(z_i)$ and ${\cal K}_2(z_i)$ are the modified Bessel  
functions  of the second kind of order $1$ and $2$.

We have  solved Eqs.~(\ref{ec_N_adim}) and (\ref{ec_mu}) in the case that the  
lepton asymmetry is
generated  by $N_1$, the lightest right-handed neutrino. 
The results for the $N_1$ abundance normalized to the equilibrium one  
$Y_N/Y_N^{eq}$ are shown in Fig.~\ref{Fig:N2Lepto}, left panel
(blue solid line).  
As was already noticed in \cite{hann}, at high temperatures 
the equilibration rate of the heavy neutrino is faster   
when the full Boltzmann equation is used, 
so the abundance is larger than the one obtained assuming kinetic equilibrium 
(pink dashed line). 

After conversion by sphaleron transitions,    
the resulting baryon asymmetry is related to the generated lepton asymmetry 
$\YLf$ by \cite{ht} 
\be
Y_B= \frac {12}{37} \YLf 
\ee

In Fig.~\ref{Fig:lepto_normal} we plot our results for 
the Baryon Asymmetry of the Universe (BAU) $Y_B$ 
(blue solid line) and  
the ones obtained using the integrated equations (pink dashed line)
for $K_1=1, 10$ and $\epsilon_1 =  10^{-6}$.  
We find, in agreement 
with \cite{hann}, that the difference between the two approaches is  
at most 20\% for $K_1 \gtrsim 1$. For smaller values of $K_1$
the difference can be larger, but in this limit 
our results are no longer valid because we have not included 
scattering which can enhance the neutrino production and change our results.

\FIGURE[t]{\includegraphics[width = 0.9\textwidth]{Fig/lepto_normal.eps}
        \caption[Example]{BAU obtained from the integration
       of the full Boltzmann equations, Eqs.~(\ref{ec_N_adim}) and 
       (\ref{ec_mu}), (blue solid line) and the 
       approximate integrated equations, 
       \eq{int_BESN},(\ref{int_BESL})  
       (pink dashed line) for
       $K_1=1$ (left) and $10$ (right), and $\epsilon_1 = 10^{-6}$.}%
	\label{Fig:lepto_normal}}

 Next, we focus on a different scenario: we assume that a sizeable lepton 
asymmetry has been produced initially
(for instance, during $N_2$ decay), and we study  
the washout of this asymmetry by $N_1$ interactions, relevant at 
$T \sim M_1 \ll M_2$, when the lepton asymmetry 
generated by $N_1$ decays is negligible, which is always the case if
$M_{N_1} \ll 10^9$ GeV \cite{oscar,riottodibari}.  Here,  as discussed 
in \cite{N2},  such lepton asymmetry may survive in
 two ways: (i) because $N_1$ interactions are weak, so it only washes out a 
small part of the lepton asymmetry;  
 (ii) if 
the $N_1$ decays when only the $\tau$ Yukawa interactions are in 
 equilibrium, at $10^{9}$ GeV $< T < 10^{12}$ GeV, generically a part of the 
 preexisting lepton asymmetry is always protected from $N_1$ washout  
due to flavour effects.
We first study the effect of using the full Boltzmann equations to follow  
 the evolution of the preexisting lepton asymmetry, due to $N_1$ interactions, 
in the unflavoured case.  

\FIGURE[t]{\includegraphics[width = 0.99\textwidth]{Fig/N2Lepto.eps}
  \caption[Example]{Comparison between the integration of the full
    Boltzmann equations (Eqs.~(\ref{ec_N_adim}) and (\ref{ec_mu})) in
    $N_2$ leptogenesis (blue solid line) and the approximate
    integrated equations (\eq{int_BESN},\eq{int_BESL}) (pink dashed
    line) for $K_1=0.1$, $1$, $10$, and $Y_{\cal{L}} = 10^{-9}$. In
    the plots on the left we show the heavy neutrino ($N_1$) abundance
    as a function of $z$, and in the ones on the right the surviving
    Baryon Asymmetry after $N_1$ washout.}
	\label{Fig:N2Lepto}}

\FIGURE[t]{\includegraphics[width = 0.7\textwidth]{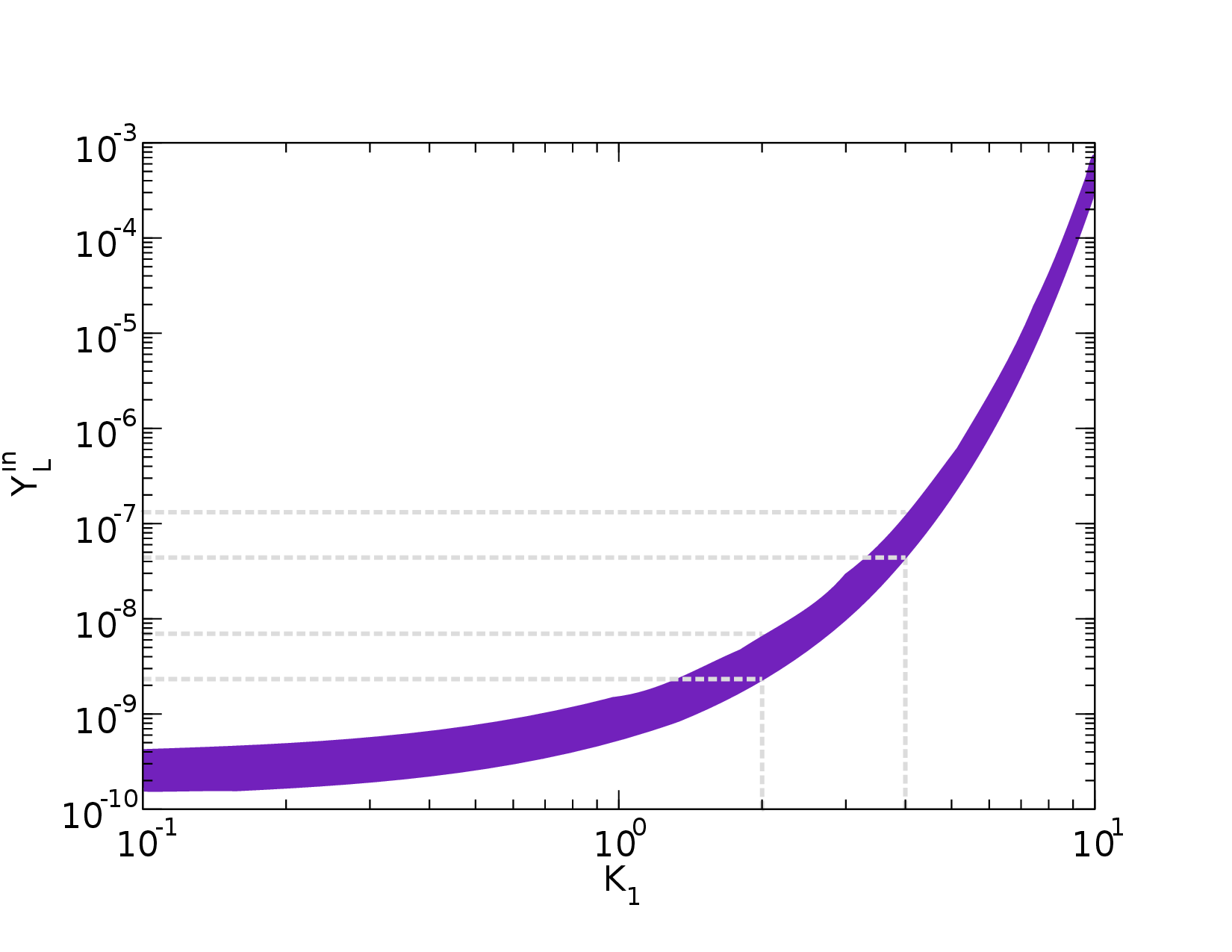}
\caption[Example]{The coloured region shows the amount of initial lepton
     asymmetry that needs to be produced by $N_2$ in order to obtain
     the right amount of BAU after $N_1$ washout, as a function of
     the washout parameter $K_1$.}
   \label{Fig:N2_contour}}

We therefore solve eqs.~(\ref{ec_N_adim}) and (\ref{ec_mu}) with the
initial conditions at $z_1 = M_1/T << 1 $ of zero $N_1$ initial
abundance and $Y_{\cal{L}} = 10^{-9}$, for different values of
the decay parameter $K_1$ and $\eps_1=0$, i.e., we neglect the lepton
asymmetry generated by $N_1$ decays.  
In this approximation, the Boltzmann equation for the lepton asymmetry
can be solved analytically and we obtain:
\be
\label{n2}
Y_{\cal{L}}(z) = Y_{\cal{L}}(0)  
\exp{\left\{ - \frac{3 K_1 }{\pi^2} 
\int_{0}^{z} dz_1 z_1^2
\int_{z_1}^{\infty} d\e_1
\int_{\frac{\e_1 - y_1}{2}}^{\frac{\e_1+y_1}{2}} dy_L
  \frac{e^{y_L}}{(1+e^{y_L})^2}  (\fHeq+f_{N_1}) 
 \right\}}
\ee
We show our results in Fig.~\ref{Fig:N2Lepto}. 
As we explained before, the $N_1$ abundance 
(left panel) is independent of the 
lepton asymmetry (to the order we are considering), so it
is the same as in the previous case.
In the right panel we show the evolution of the baryon asymmetry, as 
a function of $z=M_1/T$. We obtain  
that the use of the full Boltzmann equations always decreases the final
lepton asymmetry, as compared with the standard approximation. 
This can be understood from \eq{n2}: 
in the usual Maxwell--Bolzmann approximation, only the $\fHeq$ distribution
appears in the exponential washout factor. However in the full 
equation there is an extra term, $f_{N_i}$, and thus the
washout is stronger. We can see in this figure that for $K_1 \leq 1$ the final
asymmetry is only slightly reduced with respect to the integrated equations
and this suppression increases with $K_1$.

In Fig.~\ref{Fig:N2_contour}, we show the amount of initial lepton
asymmetry that needs to be produced by $N_2$ in order to obtain
the observed BAU after $N_1$ washout, as a function of the parameter $K_1$.
The coloured region indicates the values of $Y_L^{in}$ and $K_1$
for which we obtain  $Y_B$ within 50\% of the 
experimental  value   
$Y_B = (8.75 \pm 0.23) \times 10^{-11}$
\cite{wmap}.
In the case considered here of hierarchical
right-handed neutrinos, the asymmetry generated by $N_2$ is typically
$Y_L^{N_2} \lesssim 10^{-8}$. 
Therefore, in typical scenarios, only for $K_1 \leq
3$ we can expect a sufficient asymmetry to be generated. In fact, as shown in
\cite{riottodibari}, we usually need $K_1 \leq 3$ in $SO(10)$-inspired
scenarios with flavour effects.  

Finally,  we consider the addition of flavour effects in this scenario of an 
initial lepton asymmetry and its washout
by $N_1$ interactions. As shown in Appendix~\ref{appA}, the Boltzmann equation 
for the $N_1$ abundance is identical to the single flavour case, while now we
have to consider different evolution equations for the asymmetries in the 
different flavours:   
\bea  
 \label{ec_N_adim2} 
\frac{\partial f_{N_i}}{\partial z_i}& =& \frac{K_i z_i^2}{y_i \e_i}  
\int_{\frac{\e_i - y_i}{2}}^{\frac{\e_i + y_i}{2}}   
dy_H \left[\fHeq \fLeq (1-f_{N_i}) - f_{N_i} (1-\fLeq)(1+\fHeq)  \right] \, ,
\eea
\bea
\label{ec_Lalpha_adim}
\frac{\partial f_{{\cal L}_\alpha}}{\partial z_i} & =  & \frac {z_i^2}{y_L^2}  
\int_{y_L + \frac{z_i^2}{4 y_L}}^{\infty} d\e_i
 \left\{K_i \epi^\alpha \, (f_{N_i} - \fNieq)    
\left[(1-\fLeq)(1+\fHeq) - \fHeq \fLeq \right]
 \right. - \nonumber \\   
 &  & \left. - \frac {K_i^\alpha}{ 2}  f_{\cal{L}_\alpha} (\fHeq+f_{N_i})
 \right\} ~+~O(Y_{\alpha i}^2 Y_{\beta i}^2 \times f_{{\cal L}_\beta} )\ ,  
\eea  
where $K_i^\alpha  \equiv \Gamma (N_i \to L_\alpha H,  
\bar L_\alpha \bar H)/H(T=M_i)$. In this expression we can see that, up to
subleading corrections in the Yukawa couplings (barring special cases where 
$Y_{\beta i}^2 f_{{\cal L}_\beta} \gtrsim f_{{\cal L}_\alpha}$),  the evolution
equations for the  asymmetries
in different flavours decouple and the situation is analogous to 
the single flavour case. The only difference in this equation is the 
presence of different CP 
asymmetries in different flavours and the fact that the washout is produced 
only by the inverse decays in the relevant flavour \cite{oscar}. Therefore, 
we can extend the results obtained above in the single flavour approximation
for instance to the three flavour case
considering independently the asymmetries in the $e$, $\mu$ and $\tau$
channels with independent washout $K_1^e$, $K_1^\mu$ and $K_1^\tau$.  
Figs.~\ref{Fig:N2Lepto} and \ref{Fig:N2_contour} are also valid for the
asymmetries in the different flavours with the corresponding washout
parameter. Although our results are model independent and can be used in a
generic model, they basically agree with the results obtained in
Ref~\cite{riottodibari} in an $SO(10)$ scenario with integrated equations 
as they have always small washout in the $\tau$-flavour in the relevant
parameter space.

\section{Soft leptogenesis} 
\label{slepto}

In this section, we briefly review the soft leptogenesis scenario, and  
we present the full Boltzmann equations relevant for soft leptogenesis in the 
context of the supersymmetric type I seesaw model. We solve  
them exactly and compare with the approximations used
in the literature.  
 
For a hierarchical spectrum of right-handed neutrinos,  
successful leptogenesis requires generically quite heavy singlet neutrino  
masses~\cite{di}, of order $M>2.4 (0.4)\times 10^9$~GeV for vanishing
(thermal) initial neutrino densities~\cite{di,Mbound},  
although flavour effects \cite{flavour} and/or extended scenarios
\cite{ma} may affect this limit
\footnote{This bound applies when the lepton asymmetry is generated  
in the decay of the lightest right-handed neutrino. The possibility 
to evade the bound producing the asymmetry from the second lightest 
right-handed neutrino has been considered in \cite{db1}, and  
flavour effects have been analysed for this case in \cite{oscar}.}. 
The stability of
the hierarchy between this new scale and the electroweak one is 
natural in low-energy supersymmetry, but in the supersymmetric seesaw
scenario there is some conflict between the gravitino bound on the 
reheating temperature and the thermal production of right-handed
neutrinos \cite{gravi}.  This is so because in a high temperature 
plasma, gravitinos are copiously produced, and their late decay could
modify the light-nuclei abundances, contrary to observation. This sets 
an upper bound on the reheating temperature after inflation, $T_{RH} <
10^{6-8}$ GeV, which may be too low for the right-handed neutrinos to 
be thermally produced \cite{gravi}.

Once supersymmetry (SUSY) has been introduced, leptogenesis is induced also 
in singlet sneutrino decays.  If supersymmetry is not broken, the
order of magnitude of the asymmetry and the basic mechanism are the 
same as in the non-supersymmetric case. However, as shown in 
Refs.\cite{soft1a,soft1b,soft2}, supersymmetry-breaking terms can play an  
important role in the lepton asymmetry generated in sneutrino decays
because they induce effects which are essentially different  
from the neutrino ones.  
Soft 
supersymmetry-breaking terms involving the singlet sneutrinos remove
the mass degeneracy between the two real sneutrino states of a single 
neutrino generation, and provide new sources of lepton number and CP
violation. As a consequence, the mixing between the two sneutrino 
states generates a CP asymmetry in the decay, which can be sizable for
a certain range of parameters. In particular, the asymmetry is large 
for a right-handed neutrino mass scale relatively low, 
in the range $10^{5}-10^{8}$ GeV, below the reheating temperature limits,  
what solves the cosmological gravitino problem.  
This is the so called  ``soft leptogenesis'' scenario, which we are 
going to consider now.

The superpotential of the supersymmetric seesaw model 
contains the following relevant terms: 
\be 
\label{W} 
W = \frac{1}{2} M_{ij} N_i N_j +  Y_{\alpha i} N_i (L_\alpha  H)   \ , 
\ee
where the parentheses indicate SU(2) contractions and $\alpha, i =1,2,3$  
are flavour indices. 
$L_\alpha,N_i, H$ are the chiral 
superfields corresponding to the 
left-handed lepton doublets,   
the right-handed (RH) neutrinos and the 
up-Higgs doublet, respectively,  
and $Y_{\alpha i}$ 
denote the neutrino Yukawa couplings (notice that we are using 
the convention $\alpha\equiv L,~j \equiv R$).
The soft supersymmetry-breaking terms involving the heavy sneutrinos  
$\widetilde{N}_i$ are 
\be   
\label{soft} 
{\cal L}_{soft} =  
- \widetilde{m}^2_{ij}\widetilde{N}_i^* \widetilde{N}_j
- \left[ A_{\alpha i} Y_{\alpha i} \widetilde{N}_i (\widetilde{L}_\alpha H)   
+ \frac 1 2 B_{ij} M_{ij} \widetilde{N}_i \widetilde{N}_j + h.c.  \right] 
\ee 
Contrary to the traditional leptogenesis
scenario, where at least two generations of RH  neutrinos are 
required to generate a CP asymmetry in neutrino/sneutrino decays, in
this mechanism for leptogenesis, a single generation of heavy RH  
neutrinos is sufficient to generate a CP asymmetry in sneutrino
decays.  
Therefore, from now on we consider a simplified one-generation model,
which refers to the lightest of the three heavy sneutrinos, that we 
denote as 1.
For simplicity we also assume proportionality of soft-trilinear terms,  
and drop the flavour index for the coefficent $A$.
The sneutrino interaction Lagrangian is then: 
\begin{eqnarray}  
\label{L}
\cal{L} &=&  
-(\widetilde{m}^{2}+\left| M\right|^{2})\widetilde{N}^* \widetilde{N}
- \frac 1 2 \left( B M \widetilde{N} \widetilde{N} + h.c.\right) 
\nonumber \\
&-&   
  \left [Y_{\alpha 1} \widetilde{N} (L_\alpha h) + 
M Y_{\alpha 1} \widetilde{N}^* (\widetilde{L}_\alpha  H)  
+ A Y_{\alpha 1} 
\widetilde{N} (\widetilde{L}_\alpha  H) + h.c. \right] 
\end{eqnarray}
where $h$ is the fermionic partner of the Higgs doublet $H$. 
Under these conditions, a physical CP-violating phase is 
still present in the neutrino sector,  
\be 
\Phi = {\rm arg} (AB) \ ,
\ee 
which we choose to assign to $A$.
The right-handed neutrino has a mass $M$, while the sneutrino and 
antisneutrino states mix in the mass matrix, with mass  
eigenvectors  
\bea
\label{masseigenstates} 
\widetilde{N}_+ &=& \frac{1}{\sqrt 2} \left( e^{i \Phi/2}
\widetilde{N} + e^{-i \Phi/2} \widetilde{N}^* \right) 
\nonumber \\
\widetilde{N}_- &=& \frac{-i}{\sqrt 2} \left(e^{i \Phi/2} 
\widetilde{N} - e^{-i \Phi/2} \widetilde{N}^* \right)
\eea 
and mass eigenvalues 
\be
M_{\pm}^2 = M^2 + \widetilde{m}^{2} \pm |BM|   \ . 
\ee
We define the fermionic and scalar CP asymmetries in the decay of each 
$\widetilde{N}_i$ ($i=\pm$) as: 
\begin{eqnarray} 
\esi & = &\frac{\Nis - \Nisb}{\Nis + \Nisb} = 
 \frac{\Nis - \Nisb}{\As}
\\  
\efi & = & \frac {\Nif - \Nifb}{\Nif + \Nifb}
= \frac{\Nif - \Nifb}{\Af} 
 \label{asymdefhat} \ , 
\end{eqnarray} 
where we have implicitly summed over flavours, and
\begin{eqnarray} 
\As&=& 2 \; \sum_\alpha  |Y_{\alpha 1} M|^2 \; , 
\nonumber \\ 
\Af&=& 2 \; \sum_\alpha |Y_{\alpha 1} M|^2
\frac{M_{i}^2}{M^2} \;  (1-x_L-x_h) \; ,  
\label{amplidef} 
\end{eqnarray}
with  $M_i = M_+,M_-$ and
\be 
x_a \equiv \frac{m_a(T)^2}{M^2} \; . 
\ee
Notice that, in this section, we keep the  thermal masses of the sneutrino decay 
products, since they break supersymmetry and contribute to  
obtain a non-vanishing CP asymmetry. They are given by \cite{thermal}:
\bea 
\label{tmh} 
m^2_H(T) & = 2 m^2_h(T) & = 
\left (\frac 3 8 g_2^2 + \frac 1 8 g_Y^2 + \frac 3 4 Y_t^2 \right) T^2\; ,  
\\
\label{tml} 
m^2_{\widetilde L} (T) & = 2 m^2_L(T) & = 
\left (\frac 3 8 g_2^2 + \frac 1 8 g_Y^2 \right) T^2 \; , 
\eea 
where $g_2$ and $g_Y$ are the gauge couplings and $Y_t$ is the top Yukawa.
Then, the total sneutrino decay width is  
\footnote{Neglecting soft SUSY-breaking corrections and thermal 
masses $\Af=\As$.},    
\begin{equation}
\Gamma_i = \frac{1}{16\pi M_i} 
(\lambda(1,x_{\widetilde L},x_H)\As + 
\lambda(1,x_{L},x_h) \Af) \, , 
\end{equation} 
where 
\be  
\lambda(1,x,y) = \sqrt{(1+x-y)^2-4x}
\ee 

The mixing between the sneutrino states can generate a sizable  
CP asymmetry in their decay, due to the resonant enhancement of the 
self-energy contribution. 
The CP asymmetry can be computed 
following the effective field-theory approach described in \cite{pi},
which takes into account the CP violation 
due to mixing of nearly degenerate states by using
resumed propagators for unstable (mass eigenstate) particles.
Neglecting supersymmetry breaking in vertices
and  
keeping only the lowest order contribution in the soft terms, 
it is given by \cite{soft1a,soft2}: 
\be 
\label{epsi}
\esp = \esm = - \efp = - \efm 
\equiv \epsilon = \frac{4 \Gamma B}{\Gamma^2 + 4 B^2} 
\frac{{\rm Im} A}{M} \ . 
\ee 
Here,  
we have neglected thermal corrections to the CP asymmetry from the loops, 
i.e., we have computed the imaginary part of the one-loop graphs using  
Cutkosky cutting rules at $T=0$. These corrections 
are the same for scalar and fermionic decay channels, 
(only bosonic loops contribute to the wave-function 
renormalization common to both decays), so they are not expected to introduce  
significant changes to our results.
 
In order to generate enough asymmetry, we need $B\simeq \Gamma$, thus 
the lepton number violating soft 
bilinear coupling $B$, responsible of the sneutrino mass splitting,
has to be unconventionally small.  
Moreover, as one can see in the above equation, in soft leptogenesis 
induced by CP violation in mixing, an exact cancellation  
occurs between the asymmetry produced in the fermionic and bosonic 
channels at $T=0$. Therefore, thermal effects play a fundamental  
role in this mechanism: final-state Fermi blocking and Bose stimulating
factors, together with the effective masses of the particle excitations  
in the plasma, break supersymmetry and remove this degeneracy. 
 
Several comments are in order:

The effects of flavour in soft leptogenesis have been studied in  
\cite{concha1}, and we will not consider them here, since the
single-flavour approximation is    
enough to illustrate our results.

It has been recently pointed out that for resonant scenarios the  
use of quantum BE may be relevant \cite{QBE}.  
For {\bf standard resonant leptogenesis} \cite{resonant}, 
they induce a $T$ dependence in the
CP asymmetry which can enhance the produced baryon number.  
However, in Ref.~\cite{concha2} it has been shown that in 
{\bf soft leptogenesis}, due to the thermal nature of the 
mechanism already at the classical level, the introduction of quantum 
effects does not lead to such enhancement and therefore it is enough to  
consider only the classical BE in this work. 

The authors of Ref.~\cite{soft1b} identified new sources for soft  
leptogenesis,
induced by CP violation in right-sneutrino decay and in the interference  
of mixing and decay. These contributions are relevant both because 
they can be sizable for natural values of the $B$ parameter and  
because, unlike the CP violation in mixing, they do not require 
thermal effects, as they do not vanish at $T=0$.  
However, this calculation has been recently revisited in Ref.~\cite{concha3},
where it has been found that for all soft SUSY breaking sources  
of CP violation considered, at $T=0$ the exact cancellation between 
the asymetries produced in the fermionic and bosonic channels holds. 
Therefore it seems that the full Boltzmann equations are always required
to calculate the final lepton asymmetry generated in soft  
leptogenesis.

\subsection{Full Bolzmann equations} 
 
We assume that the sneutrinos are in a thermal  
bath with a thermalization time $\Gamma^{-1}$ shorter than the 
oscillation time, $(\Delta M)^{-1}$, therefore  
coherence is lost and we can write the Boltzmann equations for the 
mass eigenstates \eq{masseigenstates}. 
As in the previous section,   
we work in a simplified scenario
including only sneutrino decays, inverse decays and resonant scattering. 
Within this approximation, we can neglect RH neutrino interactions, since 
in soft leptogenesis only sneutrino interactions generate the lepton
asymmetry.

We assume that, because of the fast gauge interactions, 
 the Higgs 
and higgsino fields are in thermal equilibrium and the leptons and
sleptons are in kinetic equilibrium, so that the corresponding 
distributions are:
\begin{eqnarray}
f_H^{eq} =(e^{E_H/T}- 1)^{-1} &\; ,\qquad &  
f_h^{eq}  = (e^{E_h/T}+1)^{-1} \\
f_{L} =\displaystyle{\frac{1}{\exp[(E_L - \mu_f)/T]+1}} & \; ,\qquad  & 
f_{\bar{L}}=\frac{1}{\exp[(E_L + \mu_f)/T]+1} \;,\label{fdist}  \\
f_{\widetilde{L}} = 
\displaystyle{\frac{1}{\exp[(E_{\widetilde L} - \mu_s)/T]-1}}  &\; , 
\qquad & 
f_{\widetilde{L}^\dagger} = 
\frac{1}{\exp[(E_{\widetilde{L}} + \mu_s)/T]-1} \;  .    
\end{eqnarray}
We have introduced a chemical potential for the  
leptons, $\mu_f$, and sleptons, $\mu_s$.
We are interested in the evolution of  
the sneutrino density distributions $\fsNi$ and the fermionic 
and scalar asymmetries,  
$f_{\cal{L}}= f_L -f_{\bar{L}}$, which is given by \eq{flep}
to first order in $\mu_f$, and  
\be
\label{fslep} 
\fslep = \fLt - f_{\widetilde{L}^{\dag}} = 
\frac {2 \, e^{E_L/T }}{(e^{E_L/T } - 1)^2}\frac{\mu_s}{T}
+ {\cal O}((\frac {\mu_s}{T})^3) \ . 
\ee 
We approximate $f_L +f_{\bar{L}}\simeq 2\fLeq$
and $\fLt + f_{\widetilde{L}^{\dag}} \simeq 2 \fLteq$, where: 
\be 
\fLeq  = (e^{E_L/T } + 1)^{-1} \; , \qquad \qquad
\fLteq = (e^{E_{\widetilde{L}}/T } - 1)^{-1} \, . 
\ee
In the limit $m_a(T) \ll T$,
the chemical potentials are related to 
the corresponding asymmetries by 
\footnote{Taking into account the (s)lepton thermal masses
$m_L(T) = 0.3 T$ and $m_{\widetilde{L}}(T) = 0.4 T$, 
to first order in $\mu_f,\mu_s$ we obtain 
$$\YLf=\mu_f \, 0.329 \, \frac {T^2}{s}  \; ,   
\YLs =\mu_s \,  0.541 \, \frac {T^2}{s} \; ,$$ 
which are the actual values used in our calculation, instead of 
the 1/3 and 2/3 of Eqs.~(\ref{chem_potf}) and (\ref{chem_potf}),  
respectively. We see that the lepton 
thermal mass does not change significantly the lepton asymmetry,  
but the slepton thermal mass leads to a $20 \%$ reduction of the  
slepton asymmetry, proportional to 0.54 instead of 0.67.}
\bea \label{chem_potf} 
\YLf&=&\mu_f \, \frac {T^2}{3s} + {\cal O}((\frac {\mu_f}{T})^3) \; ,  \\
\label{chem_pots} 
\YLs &=&\mu_s \,  \frac {2 T^2}{3s} + 
{\cal O}((\frac {\mu_s}{T})^3)  \ .  
\eea

Note that the masses and widths of the two sneutrino states are equal as  
long as we neglect supersymmetry breaking effects, then 
$f_{\widetilde{N}_+}=f_{\widetilde{N}_-} \equiv f_{\widetilde{N}}$,  
and we can write a unique BE for $f_{\widetilde{N}}$; 
thus, in the following, we do not write the subindex $i$.  
The total fermionic (or scalar) asymmetry is then twice the asymmetry 
generated by one 
of the two sneutrinos.

The relevant Boltzmann equations are derived in appendix \ref{appB}, 
and can be written in 
terms of the dimensionless variables $z=M/T$,
$\e_a=E_a/T$ and $y_a=p_a/T$
($a=\widetilde{N},h,L,H,\tilde{L}$) as: 
\bea 
\label{ec_snu} 
\frac{\partial f_{\widetilde{N}}}{\partial z}& =&  
\frac{z^2}{2 y_N \e_N}  \left\{
K_f \,  
\int_{\e_h^m}^{\e_h^M} 
d\e_h
 \left[ \fheq \fLeq (1+\fsN) - \fsN (1-\fLeq)(1-\fheq)
  \right] +
\right. \nonumber \\
& & + \left.
K_s \, \int_{\e_H^m}^{\e_H^M}  d\e_H
\left[ \fHeq \fLteq (1+\fsN) - \fsN
    (1+\fLteq)(1+\fHeq) \right] \right\} \, , \label{BE_soft_snu}
\\
\frac{\partial \flep}{\partial z}  & = &  
\frac{ K_f z^2}{y_L \e_L} 
 \, \int_{\e_N^f}^{\infty} d\e_N 
 \left\{ - \epsilon (\fsN -\fsNeq)\left[ (1-\fLeq)(1-\fheq) +
    \fheq \fLeq \right] \right. - \label{BE_soft_F}
\nonumber \\
&  & \left. -  \frac 1 2 \flep (\fsN + \fheq)\right\}  
+ \frac{1}{Hz} S_g 
\, , \\
\frac{\partial \fslep}{\partial z} & = &  \frac{K_s z^2}{y_L \e_L} 
\int_{\e_N^{s}}^{\infty} d\e_N
\left\{ \epsilon (\fsN -\fsNeq)\left[ (1+\fLteq)(1+\fHeq) +
    \fHeq \fLteq \right] +\right.
\nonumber \\
&  & + \left.  \frac 1 2 \fslep (\fsNi - \fHeq)\right\} 
+ \frac{1}{Hz} \widetilde{S}_g 
\, , \label{BE_soft_S}
\eea
where  
\be
\fsNeq = (e^{E_N/T} - 1)^{-1} \; , 
\ee
and we have used the notation 
\be
K_f =  (1-x_L-x_h) \lambda(1,x_{L},x_h) K
\; , \qquad
K_s = \lambda(1,x_{L},x_h) K \; ,
\ee
with $K= \Gamma^0/H(T=M)$ defined in terms of the sneutrino decay width 
without thermal masses, 
\be
\Gamma^0 = \frac{M}{4 \pi} \sum_\alpha |Y_{\alpha i}|^2 \; .  
\ee

Since the thermal masses are different for the  final states $h L$ and 
$H\widetilde{L}$,
so are the integration limits, given by:
\be
\e_{h}^{M,m} = \frac 1 2 \left \{ E_i (1-x_{L} + x_{h}) 
\pm y_i \, 
\lambda(1,x_{L},x_{h}) \right \} \; ,
\ee
\be
\e_i^{f} = \frac{\e_L + \frac{z^2_i}{4 y_L} (1+x_L-x_h) \lambda(1,x_L,x_h)}
{\frac{(1+x_L-x_h) }{2} + \frac{\e_L}{2y_L}  \,
\lambda(1,x_L,x_h)} \; , 
\ee
and analogously for $\e_{H}^{M,m}$ and $\e_i^{s}$, just replacing $L,h$ with 
$\widetilde{L},H$ in the above equations.

The terms $S_g$ and $\widetilde{S}_g$ in Eqs.~(\ref{BE_soft_F})
and (\ref{BE_soft_S})
represent the fast gaugino 
interactions, defined in Eq.~(\ref{sg}). These 
interactions are in equilibrium and mediate processes that  
transform leptons into scalar leptons and viceversa ($L + L 
\leftrightarrow  \widetilde{L} + \widetilde{L}$). 
Thus we shall impose that $\mu_f =  \mu_s$.

As in the previous section, decays, inverse
decays, and on-shell scattering processes must be considered in order
to obtain the appropiate out-of-equilibrium condition (see 
appendix \ref{appB}).
Using the approximated distributions (\ref{flep}) and (\ref{fslep})
for $\flep$ and $\fslep$, respectively,
and integrating over the dimensionless (s)lepton momentum, $y_L$, 
we obtain the following Boltzmann equations for the 
chemical potentials, defined in Eqs.~(\ref{chem_potf}) and (\ref{chem_pots}):
\bea 
\frac 1 T \frac{d\mu_f}{dz} &= &\frac{3 K_f z^2}{\pi^2} 
\int_z^\infty d \e_N 
\int_{\e_L^m}^{\e_L^M} d\e_L
 \left\{   - \frac{\mu_f}{T} \frac{e^{\e_L}}{(e^{\e_L} + 1)^2}  
(f_{\widetilde{N}} +\fheq) -
 \right. \nonumber \\ 
 & & -  \left. 
 \epsilon \, (f_{\widetilde{N}} - f_{\widetilde{N}}^{eq})  
\left[(1-\fLeq)(1-\fheq) + \fLeq \fheq
 \right] \right\} 
-(\mu_f -\mu_s) \gamma_{g}
\, , \label{ec_muf} \\
& & \nonumber \\
\frac 1 T \frac{d\mu_s}{dz} &= &\frac{3 K_s z^2}{2 \pi^2} 
\int_z^\infty d \e_N 
\int_{\e_L^m}^{\e_L^M} d\e_L
 \left\{   \frac{\mu_s}{T} \frac{e^{\e_L}}{(e^{\e_L} -1)^2}  
(f_{\widetilde{N}} - \fHeq) + \right. \nonumber \\ 
 & & +  \left. 
 \epsilon \, ( f_{\widetilde{N}} - f_{\widetilde{N}}^{eq})  
\left[(1+\fLteq)(1+\fHeq) + \fLteq \fHeq
 \right] \right\} 
+(\mu_f -\mu_s) \widetilde{\gamma}_{g}
\, . \label{ec_mus} 
\eea
where 
we have summed over the two
sneutrino states and the integration limits are:
\be
\label{limits}
\e_{L}^{M,m} = \frac 1 2 
\left \{ \e_N (1+x_{L} - x_{h}) 
\pm y_N \, 
\lambda(1,x_{L},x_{h}) \right \} \; ,
\ee
and analogously for $\e_{\widetilde{L}}^{M,m}$, just replacing $L,h$ with 
$\widetilde{L},H$ in the above equation.
The $\gamma_g$, $\widetilde{\gamma}_{g}$ 
terms in Eqs.~(\ref{ec_muf}) and (\ref{ec_mus})
represent gaugino mediated processes $L + L 
\leftrightarrow  \widetilde{L} + \widetilde{L}$.

Schematically, we can write the equations for the chemical potentials 
as:
\bea 
\frac{d\mu_f}{dz} &= & \epsilon A + C \mu_f - (\mu_f -\mu_s) \gamma_{g}
T\ ,
\\
\frac{d\mu_s}{dz} &= & \epsilon B + D \mu_s + 
(\mu_f -\mu_s) \widetilde{\gamma}_{g} T
\ ,
\eea
while the lepton and slepton number asymmetries are related to the 
chemical potentials by 
\be
\YLf = \alpha_f \mu_f
\qquad \YLs = \alpha_s \mu_s \ .
\ee
Therefore, the evolution equation for the total lepton asymmetry, 
$\YLT = \YLf + \YLs$, is 
\be 
\frac{d\YLT}{dz} = \epsilon \left(\alpha_f A + \alpha_s B \right)
+ \alpha_f C \mu_f + \alpha_s D \mu_s  
\ee
Now, 
in order to take into account that at shorter time intervals the fast 
gaugino interactions make $\mu_s=\mu_f$, we replace this equality 
in the right-hand side of the above equation, obtaining
\be
\label{ec_YLT}
\frac{d\YLT}{dz} = \epsilon \left(\alpha_f A + \alpha_s B \right)
+ \frac{\alpha_f C  + \alpha_s D}{\alpha_f+\alpha_s} \YLT  
\ee
We have solved numerically Eqs.~(\ref{ec_snu}) and  (\ref{ec_YLT}).
Following previous approaches \cite{soft2},\cite{concha1} we
neglect the thermal masses in the evolution equation for the 
sneutrino distribution, \eq{ec_snu}, so the decay channels 
are always open. If one keeps the thermal masses, there is a
range of $T$ for which the decays 
$\widetilde{N} \rightarrow  hL, H\widetilde{L}$ are kinematically 
forbidden \cite{thermal}.
However we have checked that for the relevant values 
of $z$, the effect of the thermal 
masses in the sneutrino distribution is negligible. 

It is important, though, to keep the thermal masses in the evolution 
equation for $\YLT$, both, because they contribute 
significantly to the asymmetry and because the Bose--Einstein 
distribution is divergent at low values of $z$ for massless scalars, 
leading to an unphysical enhancement of the slepton number asymmetry.
In this case, one should also consider the 
CP asymmetry produced in the decays 
$H \rightarrow \widetilde{N}_i \widetilde{L}$ and 
$h \rightarrow \widetilde{N}_i L$, allowed at higher temperatures. 
In \cite{thermal}, this contribution has been found to be negligible 
in standard leptogenesis, so we assume that this is also the 
case in soft leptogenesis.

\subsection{Approximated Boltzmann Equations}

The standard integrated equations, \eq{int_BESN} and  \eq{int_BESL},
can not be used in the case of soft leptogenesis,
since the CP asymmetry produced in fermionic decays is exactly canceled by
the one produced in the scalar channel if Maxwell--Boltzmann statistics is
assumed. However one can find some approximate solutions in the
literature which try to estimate the baryon asymmetry generated in soft
leptogenesis.

One possibility, which was used in 
\cite{soft2} and \cite{concha1}, is to
neglect all the thermal corrections except for the ones that are crucial 
to get a non vanishing CP asymmetry, which are evaluated in the 
approximation of decay at rest of the heavy sneutrinos. 
Then, 
the Boltzmann equations are integrated in the standard way, i.e., 
assuming kinetic equilibrium and Maxwell--Boltzmann statistics
for all the particles in the plasma, obtaining:
\bea
\frac{d Y_{\widetilde{N}}}{dz} &=& - K z \, 
(Y_{\widetilde{N}}-Y_{\widetilde{N}}^{eq}) 
\frac{{\cal K}_1(z)}{{\cal K}_2(z)} \, , 
\label{int_epsT_snu} \\
\frac{d Y_{{\cal{L}}_T}}{dz} & = &2\, \epsilon(T) \, K z
(Y_{\widetilde{N}}-Y_{\widetilde{N}}^{eq}) 
\frac{{\cal K}_1(z)}{{\cal K}_2(z)} - 
\frac{K z^3}{4} {\cal K}_1(z) Y_{\cal{L}_T} \, ,
\label{int_epsT_final}
\eea
where $Y_{{\cal{L}}_T} = Y_{\cal{L}} + Y_{\tilde{\cal{L}}}$, and we have
summed over the two sneutrino states. 

The effective, temperature-dependent CP asymmetry, $\epsilon(T)$, 
includes the statistical factors and thermal corrections which 
are different  for fermions and scalars, and it is defined as:
\be \label{epsT}
\epsilon (T) =  \epsilon \, \frac{c_B-c_F}{c_B + c_F}\, ,
\ee
where 
\bea
c_B& =& \lambda(1,x_{\widetilde{L}},x_H)
 [1 + f_B(\e_{\widetilde{L}})] [1 + f_B(\e_H)]
 \; ,
 \\
 c_F &=&  (1-x_L-x_h) \lambda(1,x_{L},x_h)
 [1 - f_F(\e_L )]  [1 - f_F(\e_h)] \; ,
 \eea  
   being
$f_B$ and $f_F$ the Bose--Einstein and Fermi--Dirac distributions, 
respectively, and 
\be
\e_{\widetilde{L},H}=
\frac z 2 (1+x_{\widetilde{L},H}-x_{H,\widetilde{L}})  
\; , \qquad
\e_{L,h}= \frac z 2 (1+x_{L,h}-x_{h,L})
\ee

We expect this approximation to be accurate in the strong washout 
regime, $K \gg 1$, since in this case the kinetic equilibrium for the 
heavy sneutrino is a good approximation and, moreover,  
the final lepton asymmetry  is independent of the initial conditions, and fixed
only by the late time $z > 1$ (low temperature) evolution, when 
the thermal motion of the heavy sneutrino can be neglected.
We discuss the details in Sec. \ref{dis}.

We have also explored a different approach, based on Ref.~\cite{cs}:
we again
assume kinetic equilibrium for all the particles in the plasma, 
as well as
Maxwell--Boltzmann statistics for the heavy sneutrino,
but we keep the phase space and statistical factors 
which are crucial in soft leptogenesis.
Furthermore, we approximate the Fermi and Bose distributions by
the Maxwell--Boltzmann one:
\be
f_{F(B)} = \frac{1}{e^{E/T} \pm 1} \simeq e^{-E/T} \, ,
\ee
since this is enough to have a non vanishing total CP asymmetry
in sneutrino decay, and the neglected 
terms  are further suppressed by extra 
${\cal O}(e^{-E/T})$ factors.

Taking into account the (s)lepton thermal masses, 
given in \eq{tml}, the relation between the (s)lepton asymmetry
$\YLf$ and the corresponding chemical potential 
is now 
\be
\YLf = \mu_f \frac{2  m_L^2}{s \pi^2} {\cal K}_2(m_L/T) 
+ {\cal O}((\frac{\mu_f}{T})^3) \; , 
\ee
and the same expression holds for $\YLs$, just changing
$\mu_f \rightarrow \mu_s$ and $L \rightarrow \widetilde{L}$. 

Within these approximations, the Boltzmann equation for the heavy-sneutrino
abundance is again the standard one, \eq{int_epsT_snu}, 
while the 
evolution equations for the (s)lepton asimmetries become:
\bea
\label{Ylep}
\frac{d\YLf}{dz} &=& \frac{K_f z^2}{ 2\pi^2} 
\int_z^\infty d \e_N 
\int_{\e_L^m}^{\e_L^M} d\e_L
 \left\{   - \frac{\pi^2}{2} \frac {\YLf \, T^2}{m_L^2 {\cal K}_2(m_L/T)}  
\fLeq \fheq 
 \right. \nonumber \\ 
 & & 
  -  \left. 
 \eps \, (f_{\widetilde{N}} - f_{\widetilde{N}}^{eq})  
\left[(1-\fLeq)(1-\fheq) + \fLeq \fheq
 \right] /s\right\} \, , \\
& & \nonumber \\
\label{Yslep}
\frac{d\YLs}{dz} &= & \frac{K_s z^2}{ 2\pi^2} 
\int_z^\infty d \e_N 
\int_{\e_L^m}^{\e_L^M} d\e_L \left\{ -  \frac{\pi^2}{2} 
\frac {\YLs \, T^2}{m_{\sL}^2 {\cal K}_2(m_{\sL}/T)} 
\fLteq \fHeq 
 \right. \, , \nonumber \\ 
 & & +  \left. 
 \eps \, (f_{\widetilde{N}} - f_{\widetilde{N}}^{eq})  
\left[(1+\fLteq)(1+\fHeq) + \fLteq \fHeq
 \right] /s\right\} \, . 
\eea
The integration limits $\e_{L,\widetilde{L}}^{m,M}$ are given 
in \eq{limits}. 
Recall that fast gauge interactions imply that 
$\mu_f =  \mu_s$.

The integration over the (s)lepton energy can be performed 
analytically, and by writting the result as a series expansion 
in the heavy-sneutrino momentum, $y_N$, it is also possible to 
compute the integral over $\e_N$, order by order in $y_N$.
The calculation is lengthy but straightforward, and it is described in  
appendix~\ref{appC}. 
The integrated Boltzmann equation for the total lepton asymmetry
$Y_{{\cal{L}}_T}$ can be written as:
\be
\label{int_JN_final}
\frac{d Y_{{\cal{L}}_T}}{dz} = 2 \, \epsilon \, K 
(Y_{\widetilde{N}}-Y_{\widetilde{N}}^{eq}) \frac{F_1(z)}{{\cal K}_2(z)} - 
\frac{K  z^3}{4} {\cal K}_1(z) F_2(z)Y_{\cal{L}_T} \, . 
\ee
$F_1(z)$ is a complicated function of the thermal masses
that can be found in appendix~\ref{appC}. We present here the limit 
of vanishing thermal masses, for the purpose of comparison with 
Ref.~\cite{cs}:
\bea
\label{F0}
F_1^{(0)}(z) &=& 4  \sum_{n=0}^\infty  \frac{z^n}{3^{n+1} 2^{2n-1} n!} 
{\cal K}_{n+1}(3z/2)
\nonumber \\
&\simeq& \frac 8 3 {\cal K}_1(3z/2) + \frac{2z}{9} {\cal K}_2(3z/2) + 
\frac{z}{36} {\cal K}_3(3z/2) + \ldots \; ,
\eea
where the dots stand for negligible contributions. The approximation used 
in \cite{cs} seems to correspond to just keeping  the first term in 
\eq{F0}, although these authors obtain that the argument of the 
Bessel function ${\cal K}_1$ is $\sqrt{2}z$ instead of $3z/2$.
We find that to keep only the first term in $F_1(z)$ is not enough 
to get an accurate result.

The function $F_2(z)$ is given by 
\be
\label{F2}
F_2(z) =  2 T^2 \, \frac{ [\lambda(1,x_{L},x_h)]^2 (1-x_L-x_h) +
 [\lambda(1,x_{\sL},x_H)]^2}
{m_L^2  {\cal K}_2(m_L/T) + m_{\sL}^2 {\cal K}_2(m_{\sL}/T)}    \; . 
\ee
It is easy to see that it reduces to the standard washout term in the limit 
of zero thermal masses, since  
${\cal K}_2(m/T) \simeq 2 T^2/m^2$ for $m/T \ll 1$.

We expect that this approximation works better than the effective 
CP asymmetry $\epsilon(T)$, since here the thermal motion of the 
sneutrino is fully accounted for. In fact, we have checked that 
our approximated equation (\ref{int_JN_final}) 
with the first three terms of $F_1(z)$ gives the same 
results as the numerical integration, when one assumes kinetic 
equilibrium for the heavy sneutrino and Maxwell--Boltzmann statistics
for all the particles. 

\FIGURE[t]{\includegraphics[width = 0.99\textwidth]{Fig/Soft_K.eps}
\caption[Example]{Integration of the full Boltzmann
    equations in soft leptogenesis and the approximate 
    equations, for $K = 0.1$, $1$, $10$, and
    $\epsilon=4.4\cdot 10^{-6}$. Left panel: sneutrino
    distribution as a function of $z$, obtained from the integration
    of \eq{ec_snu} and the approximate \eq{int_epsT_snu}. 
    Right panel: baryon asymmetry obtained from the
    integration of Eqs.~(\ref{ec_muf}) and (\ref{ec_mus}) and the generated in 
    the two approximate
    equations \eq{int_JN_final} and \eq{int_epsT_final}.}
  \label{Fig:Soft_lepto}}

\subsection{Results and discussion}
\label{dis}

\FIGURE[t]{\includegraphics[width = 0.9\textwidth]{Fig/fkeq_f_K2.eps}
\caption[Example]{Sneutrino momentum distribution assuming 
      kinetic equilibrium (left) and 
      solving the full Boltzmann equation (right),  
      for different values of $z$, and $K = 0.1$, $1$,
      $10$.}
  \label{Fig:snu_distr_noNorm}}

Our main results in the comparison between the three approaches to the
Boltzmann equations in soft
leptogenesis are summarized in Fig.~\ref{Fig:Soft_lepto}
for different values of $K$.
In the left panel, 
we plot the sneutrino number density normalized to
the equilibrium density, from the full (solid blue) and 
integrated (dashed pink) Boltzmann equations.
In the right panel, we show 
the baryon asymmetry  obtained by integrating the 
full set of Boltzmann equations, 
Eqs.~(\ref{BE_soft_snu}) and (\ref{ec_YLT}),  
(solid blue),  
our approximated 
equations, \eq{int_JN_final} (dashed pink),
and the effective $\epsilon(T)$ approximation, \eq{int_epsT_final}, 
(dotted light blue).
Recall that in the MSSM, the final baryon asymmetry induced by
sphaleron transitions is 
\be
Y_B =\frac{8}{23} Y_{{\cal{L}}_T} \; .
\ee

We find that at high temperatures (small $z$), the equilibration rate of
the sneutrino number density  
$n_{\widetilde{N}}$ is 
higher when the full Boltzmann equations are used, 
similarly to what we have in Figure~\ref{Fig:N2Lepto} in the case of the
heavy neutrino.
The main reason for this effect is that when $z < 1$, the low momentum states 
of the sneutrino are populated very efficiently, while the population of high 
momentum modes is small and comparable to the kinetic equilibrium
distribution. This can be 
seen in Fig.~\ref{Fig:snu_distr_noNorm}, where we plot the 
sneutrino phase-space distribution, $\fsN$, assuming kinetic
equilibrium (left panel) and using the full Boltzmann equations
(right panel), 
for different values of $z$.
{}From these plots, it is clear that, 
even for $z>1$ in the weak washout, the density of low momentum 
states is significantly larger than the density obtained assuming 
kinetic equilibrium.

The behaviour of the baryon asymmetry, shown in the right panel of
Fig.~\ref{Fig:Soft_lepto}, results from the competition
of different effects.  In soft leptogenesis, due to the inclusion of
the thermal masses, the CP asymmetry vanishes for $z \lesssim 0.8$,
because both, fermionic and bosonic sneutrino decay channels are
kinematically forbiden.  For $z \lesssim 1.2$ the fermionic channel of
sneutrino (inverse) decay creates an asymmetry, which tends to flip
sign when the bosonic channel is open ($z \gtrsim 1.2$), since this
one dominates.  On the other hand (as it happens also in the standard
leptogenesis case), the asymmetry generated during the $\widetilde{N}$
production has opposite sign (we call it ``wrong-sign'' asymmetry) to
the one produced at later times, in $\widetilde{N}$ decay
(``right-sign'' asymmetry).  In the different panels of
Fig.~\ref{Fig:Soft_lepto} one can see the competition of these two
effects, depending on the strength of the Yukawa sneutrino
interactions.

In the case $K=0.1$, we observe (in agreement with 
\cite{soft2}) the flipping of the lepton asymmetry sign when 
the bosonic channel is open, both in the full (solid blue) and 
in the approximated calculations. 
Then, since the Yukawa interactions are 
weak and the decay occurs at late time, the generation of the 
right-sign asymmetry is not enough to overcome the wrong-sign 
one, created during the sneutrino production. 
However there is an important difference between assuming or not 
kinetic equilibrium for the heavy sneutrino. 
As can be seen in the first row of Fig.~\ref{Fig:snu_distr_noNorm}, 
the sneutrino momentum distribution obtained with the full equations  and for 
all values of $z$ is always larger than the
kinetic-equilibrium distribution. 
{}From Eqs.~(\ref{BE_soft_F}) and (\ref{BE_soft_S}), 
we can see that the source of the asymmetry is
always proportional to $f_{\tilde N} -f_{\tilde N}^{eq} $, where 
$f_{\tilde N}^{eq} $ is the Bose--Einstein distribution. 
In Figure \ref{Fig:snu_distr_Norm}, we can see that up to $z\lesssim 1$, 
$f_{\tilde N} -f_{\tilde N}^{eq} \simeq -f_{\tilde N}^{eq} $ 
while for $z\gtrsim 1$, 
$f_{\tilde N} -f_{\tilde N}^{eq} \geq 0$ for most values of the momenta. 
As a consecuence, the right-sign asymmetry for $z\gtrsim 1$ is larger 
than in the kinetic-equilibrium
case, partially compensating the wrong-sign asymmetry generated in the
production although it is still not big enough  
to flip the sign again. 
Notice that this is not the usual washout, which is very small  
for $K= 0.1$.
If one assumes kinetic equilibrium, the population of 
the momentum states is more uniform, and the right-sign 
asymmetry generated is smaller, so one gets a larger  
final result. 
However for $K < 1$  one has to be cautious, since  
scattering processes, that we have not included,  
can be relevant (see for instance \cite{nrr}).

For $K=1$, we see in the second row of Fig.~\ref{Fig:Soft_lepto} that 
assuming kinetic-equilibrium distributions we find two  
sign-flips: one due to the opening of the bosonic decay channel and
the second, at later time, because for larger $K$, the 
washout of the initial wrong-sign asymmetry is more efficient, 
and an asymmetry of the right-sign is finally created. 
However, using the full Boltzmann equation for the sneutrino, 
these two effects partially compensate each other. In fact, the
bosonic channel is open at $z \gtrsim 1.2$ and, as we see in Figure
\ref{Fig:snu_distr_Norm}, $f_{\tilde N} -f_{\tilde N}^{eq} \geq 0$ precisely
for $1 \lesssim z \lesssim 3$. 
Therefore, both changes of sign occur at the same time
and there is no sign-flip at all. 
The final baryon asymmetry using the full Boltzmann equations 
is in this case two (four) times bigger 
than our (the effective $\epsilon(T)$) approximation. 
For these values of $K$, the thermal motion of the sneutrino
is important, so our analytic approximation gives a better 
agreement, specially at high $T$.

Finally, for $K=10$, we find no  sign-flip of the baryon asymmetry in the 
full Boltzmann equation calculation,  
nor in the approximated ones, since  
 the right-sign asymmetry generated in the sneutrino 
decay (which occurs at earlier times) compensates the  
sign-flip due to the opening of the bosonic decay 
channel.
As we anticipated, in this strong washout scenario 
both approximations are very good, 
since, in this regime, kinetic equilibrium is rapidly reached 
(Fig.~\ref{Fig:snu_distr_noNorm}, third row), 
and the thermal motion of the sneutrino is negligible
for the relevant values of $z$, so the effective $\epsilon(T)$
approximation gives also accurate results.

\FIGURE[t]{\includegraphics[width = 0.9\textwidth]{Fig/ffeq_K.eps}
\caption[Example]{Sneutrino momentum distribution normalized to the
      equilibrium one, the Bose--Einstein distribution. We show the
      ratio $f_{\tilde{N}}/f_{\tilde{N}}^{eq}$ for different values
      of $z$ and $K = 0.1$ (left), $K = 1$ (right).}
  \label{Fig:snu_distr_Norm}}


\section{Conclusions}

The integrated
Boltzmann equations are generally used to estimate the produced
baryon asymmetry through the leptogenesis mechanism; however, these
equations are not always a good approximation. In \cite{hann}, it was
shown that in the strong wash-out regime the results may differ by
$15-30\%$. In our work, we have extended this study to two different
scenarios where the use of the full Boltzmann equations may be
relevant: the washout of a preexisting asymmetry (for instance 
generated by a heavier right handed neutrino, $N_2$)
and soft leptogenesis.

We consider a simplified picture which  
includes only the heavy right-handed neutrino ($N_i$)
decays, inverse decays, and  
resonant scattering. This approximation is appropriate
 for $T<10^{12}$ GeV, when 
the off-shell $2 \leftrightarrow 2$ scattering
mediated by $N_i$ has only small effects.

A lepton asymmetry produced during the next-to-lightest right-handed neutrino 
($N_2$) decay, in general is washed-out later by the lightest singlet neutrino 
($N_1$) interactions. Then one wonders if a sufficient 
asymmetry can survive to explain the observed baryon asymmetry. 
Our results show that the wash-out obtained using the full
Boltzmann equations is  stronger than the one obtained in
the usual Maxwell--Boltzmann and kinetic equilibrium approximation.
This result remains valid when flavour effects and different
flavour asymmetries are taken into account.

The second scenario that we have explored is soft leptogenesis, where
the lepton asymmetry is produced by sneutrino interactions, and the source of
lepton number and CP violation are the soft SUSY breaking parameters. 
The lepton asymmetry produced within this mechanism is a pure 
thermal effect, because at $T=0$ the asymmetry in leptons cancels the one in 
sleptons, and only when $T \neq 0$, the different statistics 
of bosons and fermions lead to a non-zero asymmetry.
In this context, the use of the full Boltzmann equations is mandatory, 
since in the usual Maxwell--Boltzmann approximation the produced lepton
asymmetry is exactly zero.
However, there are various simplified approaches 
available in the literature, always assuming kinetic equilibrium for the 
heavy sneutrino. A possibility is to use an effective, temperature 
dependent CP asymmetry, $\epsilon(T)$, which captures the main 
thermal effects but neglects the thermal motion of the heavy sneutrino
\cite{soft2}. 
Another approach is to expand the Bose--Einstein and Fermi--Dirac
distributions about the Maxwell--Boltzmann one, keeping only
the leading order terms \cite{cs}.
We have found an improved approximation, Eq.~(\ref{int_JN_final}),
which describes accurately
the temperature dependence of the CP asymmetry, including the thermal
motion of the sneutrino. 
We have compared our results from the full Boltzmann equations with these 
approximate solutions.

We find that, in the strong washout regime ($K \gg 1$), 
the results obtained using the full set of Boltzmann equations and
those obtained with the approximate equations are in very good agreement,
because in both cases the heavy-species distribution is very close 
to equilibrium in the relevant range of $z$. 
However, in the weak washout regime ($K \lesssim 1$), there are
important effects. Using the full kinetic equations, we find that 
the heavy-species low-momentum modes are more efficiently populated, 
leading to a lepton asymmetry that can be either enhanced (as in 
the case $K=1$) or suppressed ($K=0.1$) with respect to the 
values obtained using the approximate equations. 
In fact, the difference bewteen the full kinetic equations 
and the different approximations can be up to one order of
magnitude.

\section*{Note added}
After the present paper was submitted, a new analysis of standard
leptogensis with the full Boltzmann equations appeared \cite{plu}. The
authors of this work extend previous studies by including, in addition
to decays and inverse decays, scattering processes of the right-handed
neutrino with the top quark.

\section*{Acknowledgments}
We would like to thank M.C. Gonz\'alez-Garc\'\i a  for useful discussions.  
J.G. is supported by a MICINN-FPU Spanish grant.
This work is supported in part by the Spanish MICINN grants 
FPA-2007-60323,  FPA2008-02878 and FPA2008-00319,
and the Consolider Ingenio-2010 project 
CSD2007-00060, by the European Commission (RTN network MRTN-CT-2004-503369 and
the ILIAS project RII3-CT-2004- 506222),
and by Generalitat Valenciana under grants
PROMETEO/2008/004, PROMETEO/2009/116 and PROMETEO/2009/091.

\appendix

\section{Full BEs in standard leptogenesis}
\label{appA}

In this appendix, we describe in detail the derivation of 
the full Boltzmann equations
(BEs) relevant for standard leptogenesis, including 
the subtraction of the 
on-shell two-body scattering amplitude which leads to the 
required Sakharov condition that no lepton asymmetry can 
be generated in thermal equilibrium.

The Boltzmann equation for the phase-space distribution of the particle  
species $a$ can be written as:
\be \label{bol1}
\frac{\partial f_a}{\partial t} - H {p} 
\frac{\partial f_a}{\partial{p}} = -\frac{1}{2E} C[f_a] \, ,
\ee
where the collision integral is given by: 
\bea
C[f_a] & = & \sum_{aX \leftrightarrow Y} \int d\Pi_X d\Pi_Y (2 \pi)^4 
\delta^4 (p_a + p_X - p_Y) \nonumber \\
 &  \times & \left[ f_a f_X (1 \pm f_Y) |A(aX \to Y)|^2   - 
f_Y(1 \pm f_a)(1 \pm f_X)|A(Y \to a X)|^2 \right] \ .
\eea
The sum runs over all allowed processes $aX \leftrightarrow Y$, where
$X$ and $Y$ are multiparticle states, and we have used the 
abbreviations:
\bea
d \Pi_{X,Y}&=&\prod_{b \in X,Y} \frac{d^3 p_b}{(2\pi)^3 2 E_b} \; ,
\\ 
f_{X,Y} &=&  \prod_{b \in X,Y} f_b \; , \qquad \qquad
(1 \pm f_{X,Y}) = \prod_{b \in X,Y} (1 \pm f_b) \; ,
\eea
where $+$ is for bosons and $-$ for fermions. 

Then, for the heavy neutrino we have:
\bea
\label{RneutrA1}
\frac{\partial f_{N_i}}{\partial t}&-& p_{N_i} H 
\frac{\partial f_{N_i}}{\partial {p}_{N_i}} = \frac{1}{2E_{N_i}} 
\int d\vec{p}_L d\vec{p}_H 
(2\pi)^4 \delta^{(4)}(\pN - p_L - p_H)  \\
 & \times & \left\{f_H \fL (1-f_{N_i})  \ALHN 
- f_{N_i} (1-\fL)(1+f_H) \ANLH \right.\nonumber \\
&+&\left. f_{\bar H} f_{\bar L} (1-f_{N_i})  \ALHNb 
- f_{N_i} (1-f_{\bar L})(1+f_{\bar H}) \ANLHb \right \}  \ , \nonumber 
\eea
where $d\vec{p}_X \equiv d^3 p_X/2E_X (2 \pi)^3 $.
Using CPT invariance, we can write the neutrino decay amplitudes 
in terms of 
the CP conserving total decay amplitude $\ADi$ and the CP asymmetry
$\epi \ll 1$:
\begin{eqnarray}
\label{amplid}
\left|A(N_i  \to L H) \right|^2=\left|A(\bar{L} \bar{H} \to N_i)\right|^2
= \frac{1+\epi}{2}\ADi \, ,\\ 
\left|A(N_i  \to \bar{L} \bar{H})\right|^2=\left|A(L H \to N_i) \right|^2
= \frac{1-\epi}{2}\ADi \, .\nonumber  
\end{eqnarray}
We work  in the approximations of thermal equilibrium for the 
Higgs field and kinematic equilibrium for the SM leptons, as
described in Sec.~\ref{lepto}, Eqs.~(\ref{distrH}) and (\ref{distrL}). 
Then, substituting the above decay and inverse 
decay amplitudes 
and neglecting terms of order
${\cal O} (\eps f_{\cal{L}})$ 
(where $f_{\cal L} = \fL - f_{\bar L}$) and 
of order ${\cal O} ((\mu/T)^2)$
(where $\mu$ is the lepton chemical potential, see  
Eq.~(\ref{fl2})), we get
\bea
\label{bolN}
\bolN &=& 
 \frac{1}{2E_{N_i}} \int d\vec{p}_L d\vec{p}_H (2\pi)^4  
\delta^{(4)}({p}_{N_i} -{p}_L -{p}_H)  \ADi \nonumber \\
 &\times  &    
\left[ \fHeq \fLeq (1-f_{N_i}) - f_{N_i} (1-\fLeq)(1+\fHeq)  \right] \, .
\eea
We can perform part of the phase space integration and finally obtain: 
\be 
\label{EC_N}
\bolN = \frac{M_i \Gamma_i}{E_{N_i} \pN} 
\int_{\frac{E_{N_i}-\pN}{2}}^{\frac{E_{N_i}+\pN}{2}}
d\pH  \left[\fHeq \fLeq (1-f_{N_i}) - f_{N_i} (1-\fLeq)(1+\fHeq)  \right] \, .
\ee
Although we do not write the arguments of the distribution functions
for simplicity, energy conservation implies that $\fLeq = \fLeq(E_{N_i} -
E_H)$ in \eq{EC_N}.

Regarding the evolution of the lepton asymmetry, besides the right-handed 
neutrino decays and inverse decays we need to consider 
the $\Delta L=2$ scattering terms, in order to obtain the out-of-equilibrium 
condition, i.e., that no lepton asymmetry is generated in thermal equilibrium.
So we have:
\be
\label{lep}
\bollep = D_i - \bar{D}_i - 2 S \, ,
\ee
where: 
\bea
\label{D}
D_i &=& \frac{1}{2E_L} 
\int d\vec{p}_{N_i} d\vec{p}_H  
(2\pi)^4 \delta^{(4)}(\pN - p_L - p_H)  \\
 & \times & \left\{f_{N_i} (1-\fL)(1+\fHeq) \ANLH
- \fHeq \fL (1-f_{N_i})  \ALHN \right \}  \, , \nonumber 
\eea
\bea
\label{Dbar}
\bar{D}_i &=& \frac{1}{2E_L} \int 
d\vec{p}_{N_i} d\vec{p}_{\bar{H}} 
(2\pi)^4 \delta^{(4)}(\pN - p_{\bar{L}} - p_{\bar{H}})  \\
 & \times & \left\{f_{N_i} (1-\fLb) (1+\fHeq) \ANLHb
- \fHeq \fLb (1-f_{N_i})  \ALHNb \right \} \, , \nonumber 
\eea
and, 
\bea
\label{s}
S& = & \frac{1}{2E_L} 
\int  d\vec{p}_H  d\vec{p}_{\bar{L}} d\vec{p}_{\bar{H}}
(2\pi)^4 \delta^{(4)}(p_L + p_H  - p_{\bar{L}} - p_{\bar{H}}) \nonumber \\
& \times & 
      \left\{f_L \fHeq (1-f_{\bar{L}})(1+\fHeq)
|M_{sub}(L H \to \bar{L} \bar{H})|^2- \right. \nonumber \\ 
& - & \left. f_{\bar{L}} \fHeq (1-f_L)(1+\fHeq)
|M_{sub}(\bar{L} \bar{H} \to L H)|^2 \right\} \, ,   
\eea
The scattering term is defined in terms of the subtracted amplitudes, 
since the on-shell contribution is already taken into account through
the decays and inverse decays in the $D_i-terms$. So, for example: 
\begin{equation}
\label{amplisub}
\left|M_{sub}(L H \to \bar{L} \bar{H})\right|^2=
\left|M(L H \to \bar{L} \bar{H})
\right|^2-
\left| M_{os}(L H\to \bar{L} \bar{H}) \right|^2 ~,
\end{equation}
with
\begin{eqnarray}
\label{amplios}
\left| M_{os}(L H  \to \bar{L} \bar{H}) \right|^2 &=& 
\left| {A} (L H \to {N}_i)\right|^2 
\frac{\pi \delta(s-M_{i}^2)}{M_{i}  
\Gamma_i^{th}} 
\left| {A} ({N}_i \to \bar{L} \bar{H} )\right|^2 \nonumber \\
&\simeq& \frac {1 -2 \epsilon_i}{4}~ |{A}_D^i|^2 
\frac{\pi \delta(s-M_{i}^2)}{M_{i}  
\Gamma_i^{th}} 
| {A}_D^i|^2\, ,
 \end{eqnarray}  
and analogously:
\begin{eqnarray}
\label{amplios2}
\left|M_{os}(\bar{L} \bar{H} \to L H ) \right|^2 \simeq 
 \frac {1 + 2 \epsilon_i}{4}~| {A}_D^i|^2 
\frac{\pi \delta(s-M_{i}^2)}{M_{i}  
\Gamma_i^{th}} |{A}_D^i|^2  
\;.
 \end{eqnarray}  

We define also the off-shell part, which is CP-conserving at leading order, as,
\bea
\label{Moff}
|M_{off}(L H  \to \bar{L} \bar{H}) |^2 = 
|M (L H \to \bar{L} \bar H)|^2 - 
|A^i_{D}|^2 ~~{\pi \delta(s- M_i^2)\over 4 M_i \Gamma_i}~~
|A^i_{D}|^2 \, .
\eea
The width that cutoffs the resonance is $\Gamma^{th}_i$, 
the damping rate at finite temperature: 
\be
\label{gammath}
\Gamma^{th}_{i} = \frac{1}{2 M_i} \int d\vec{p}_L d\vec{p}_H 
(2\pi)^4 \delta^{(4)}(\pN - p_L - p_H) 
\left[(1-\fLeq)(1+\fHeq) + \fLeq \fHeq \right] \ADi  \, .
\ee
Using the amplitudes \eq{amplid} and the approximations in
Eqs.~(\ref{flep}) and (\ref{fl2}), we can simplify the integrand of the 
$D_i-terms$ as follows (to first order in $\epi$ and $\flep$): 
\bea \label{Dterm}
& & 
\displaystyle{\frac{\ADi}{2}} 
\big\{f_{N_i} (1-\fL)(1+\fHeq) (1 + \epi) 
- \fHeq \fL (1-f_{N_i}) (1 - \epi)\nonumber \\  
& & \quad\qquad  - f_{N_i} (1-\fLb) (1+\fHeq) (1 - \epi) 
+ \fHeq \fLb (1-f_{N_i}) (1 + \epi)  \big\} \nonumber \\
&=& \frac{\ADi}{2}  
\left\{
2 \, \epi \, \left[ (1-f_{N_i})\fHeq \fLeq + f_{N_i} (1-\fLeq)(1+\fHeq) \right]
- f_{\cal{L}} (\fHeq+f_{N_i})  \right\} \, .
\eea
Notice that if we only included the $ 1 \leftrightarrow 2$ processes,
a lepton asymmetry would be generated even in thermal equilibrium, 
when $f_{N_i} = \fNieq$ and $f_{\cal{L}} =0$, 
because the first term on the right-hand side of \eq{Dterm}  
does not vanish \cite{kolb}.
As anticipated, to remedy this problem we have to include also the 
subtracted contribution of the $ 2 \leftrightarrow 2$ scattering.
In fact, it is enough to include only the  on-shell contribution to the 
$L H \rightarrow \bar{L} \bar{H}$, as the off-shell terms will 
contribute only at higher order in the Yukawas and the asymmetry. The on-shell   
scattering is \cite{thermal}:
\bea
\label{osscat}
S_{os}(L H \rightarrow \bar{L} \bar{H}) &=&
 \frac{1}{2E_L} 
\int d\vec{p}_H d\vec{p}_{\bar{L}} d\vec{p}_{\bar{H}}
(2\pi)^4 \delta^{(4)}(p_L + p_H  - p_{\bar{L}} - p_{\bar{H}}) 
\\
&\times &  f_L \fHeq (1-f_{\bar{L}})(1+\fHeq) \ALHN 
\frac{\pi \delta(s-M_{i}^2)}{M_{i}
\Gamma^{th}_{i}} \ANLHb \, , \nonumber
\eea
Since the resonant part of  the scattering term is
 $S_{os}=S_{os}(L H \rightarrow \bar{L} \bar{H} )- 
S_{os}( \bar{L} \bar{H} \rightarrow L H )$, 
substituting  the decay and inverse decay amplitudes from \eq{amplid}
it is easy to see that it is order $\epi$.  Thus, within our linear 
approximation we can consistently use the equilibrium distributions also for 
the leptons when calculating 
this term. We then rewrite the product of equilibrium densities as: 
\be
(1-\fLeq)(1+\fHeq)  = \fNieq e^{E_{N_i}/T}
\left[(1-\fLeq)(1+\fHeq) + \fLeq \fHeq \right] ~,
\ee
and use the identity:
\be
1 = \int d^4\pN \delta^{(4)} (\pN - p_{{L}}-
p_{{H}}) ~,
\ee
to obtain $S_{os}(L H \rightarrow \bar{L} \bar{H})$ 
at the required order in $\epsilon_i$:
\bea
\label{sos1}
S_{os}(L H \rightarrow \bar{L} \bar{H}) &=&  \frac{1}{2E_L} 
\int d\vec{p}_H 
(2\pi)^4 \delta^{(4)}(\pN - p_L - p_H)
\fLeq \fHeq \left(\frac{1-\epi}{2}\right)^2 \ADi \\
&\times & 
\int \frac{d^4 \pN}{(2 \pi)^4} \frac{2 \pi
 \delta(\pN^2-M_i^2)}{2 M_i \Gamma^{th}_i} \,  \fNieq \, e^{E_{N_i}/T}
\nonumber \\
& \times & 
\int d\vec{p}_{\bar{L}} d\vec{p}_{\bar{H}}
 (2\pi)^4 \delta^{(4)} (\pN-p_{\bar{L}}-p_{\bar{H}} ) \ADi
\left[(1-\fLeq)(1+\fHeq)+ \fLeq \fHeq \right] \, .
\nonumber 
\eea
Notice that the integrals over the final-state particles reproduce the thermal
width of $N_i$, given by \eq{gammath} \cite{thermal}.
Since $\int \frac{d^4 \pN}{(2 \pi)^4} 2\pi
\delta (\pN^2-M_i^2 )=\int\frac{d^3\pN}{(2
\pi)^3 2 E_{N_i}}$, we can rewrite \eq{sos1} 
as:
\be
\label{sos2}
S_{os}(L H \rightarrow \bar{L} \bar{H}) =  \frac{1}{2E_L} 
\int d\vec{p}_H d\vec{p}_{N_i}
(2\pi)^4 \delta^{(4)}(\pN - p_L - p_H) \fNieq e^{E_{N_i}/T}
\fLeq \fHeq \left(\frac{1-\epi}{2}\right)^2 \ADi \, .
\ee
To this order, the $S_{os}( \bar{L} \bar{H} \rightarrow L H )$ 
term is the same, 
just changing the $(1-\epi)^2$ by $(1+\epi)^2$, so altogether we find
that:
\be
\label{sos3}
S_{os}=  -  \frac{\epi }{2E_L} 
\int d\vec{p}_H d\vec{p}_{N_i}
(2\pi)^4 \delta^{(4)}(\pN - p_L - p_H) \fNieq e^{E_{N_i}/T}
\fLeq \fHeq  \ADi \, .
\ee
Using  the equilibrium relation:
\be
\fNieq e^{E_{N_i}/T}
\fLeq \fHeq = \fNieq (1-\fLeq)(1+\fHeq) = (1-\fNieq) \fLeq \fHeq \ ,
\ee
we can write the Boltzmann equation for the lepton number asymmetry as:
\bea
\label{bollep}
\bollep &=& D_i - \bar{D}_i + 2 S_{os} 
 = \frac{1}{2E_L} 
\int d\vec{p}_H d\vec{p}_{N_i}
(2\pi)^4 \delta^{(4)}(\pN - p_L - p_H) 
\\ & \times& \ADi
\left \{ \epi \, (f_{N_i} - \fNieq)
\left[ (1-\fLeq)(1+\fHeq) - \fLeq \fHeq \right]  - \frac 1 2 f_{\cal{L}} 
(\fHeq+f_{N_i}) \right\}
\nonumber \\
&=& \frac{M_i\Gamma_i}{E_L \pL}\int_{\pL+\frac{M_i^2}{4\pL}}^{\infty} 
dE_{N_i} 
 \left\{   \epi \, (f_{N_i} - \fNieq)  
\left[(1-\fLeq)(1+\fHeq) - \fHeq \fLeq \right]
 \right. \nonumber \\ 
 & & \left. -\frac 1 2  f_{\cal{L}} (\fHeq+f_{N_i}) \right\} \ , 
\eea
where we have integrated over part of the phase space and we have used
that the heavy-neutrino width at zero temperature is given by 
\eq{gammai}.
This equation does not generate any lepton  asymmetry in thermal 
equilibrium, since
the right-hand side of the equation explicitly vanishes  in this case.
Notice that in this final equation for the evolution of the lepton-asymmetry 
distribution, (\ref{bollep}), we have only kept the on-shell
contribution to the $\Delta L = 2$ scattering mediated by $N_i$, 
because the off-shell part is higher order in the heavy-neutrino Yukawa 
couplings, and therefore subdominant unless these are large
\cite{review}.

In the presence of different active lepton flavours in the plasma at $T=M_i$, 
the
Boltzmann equations change. The main difference is that now the right-handed
neutrinos decay differently to the different active flavours and generate
different phase space distributions and different asymmetries in them. 
Taking this into account, it is easy to see that the equation for the abundance
of the heavy neutrino, \eq{RneutrA1}, now becomes:
\bea
\label{Rneutrfla}
\bolN = \frac{1}{2E_{N_i}} 
\int d\vec{p}_L d\vec{p}_H 
(2\pi)^4 \delta^{(4)}(\pN - p_L - p_H)\qquad\qquad\qquad \qquad~ \\
  \times  \sum_\alpha \left\{f_H f_{L_\alpha} (1-f_{N_i})  \ALHNal 
- f_{N_i} (1-f_{L_\alpha})(1+f_H) \ANLHal \right.\nonumber \\
+\left. f_{\bar H} f_{\bar L_\alpha} (1-f_{N_i})  \ALHNbal 
- f_{N_i} (1-f_{\bar L_\alpha})(1+f_{\bar H}) \ANLHbal \right \} ~~\ , 
\nonumber 
\eea
where 
\begin{eqnarray}
\label{amplidals}
\left|A(N_i  \to L_\alpha H) \right|^2=
\left|A(\bar{L}_\alpha \bar{H} \to N_i)\right|^2
= \frac{|A^i_{D, \alpha}|^2+ \epial \ADi}{2} \, ,\\ 
\left|A(N_i  \to \bar{L}_\alpha \bar{H})\right|^2=
\left|A(L_\alpha H \to N_i) \right|^2
= \frac{|A^i_{D, \alpha}|^2-\epial \ADi }{2} \, ,\nonumber  
\end{eqnarray}
being 
$A^i_{D,\alpha}$ the CP conserving decay amplitude to the $\alpha$
flavour and 
$\epsilon_i^\alpha$
the flavoured CP asymmetries. 
In the same approximations as before, 
i.e. up to terms of order $O(\epsilon f_{\cal L})$, we obtain 
for the heavy neutrino abundance the 
same equation as in the unflavoured case, \eq{bolN}.

In the evolution  equations for the lepton asymmetry, we have to consider 
asymmetries
for the different active flavours.  The equations are now:
\be
\label{lepal}
\bollepal = D_i^\alpha - \bar{D}_i^\alpha - S^\alpha +  S^{\bar \alpha}  \, ,
\ee
where $D_i ^\alpha$, $\bar D_i ^\alpha$ are identical to
Eqs.~(\ref{D}) and (\ref{Dbar}), taking into account only the relevant
flavour in the distribution functions and decay amplitudes, i.e.  $f_L
\to f_{L_\alpha}$, $\ANLH \to \ANLHal$, $\ANLHb \to\ANLHbal$, etc.

The scattering terms, in this case, are slightly complicated as the initial 
and final leptons can have now different flavours and we have additional 
scatterings changing flavour but not lepton number.
\bea
\label{salb}
S^{\alpha}& = & \frac{1}{2E_L} 
\int  d\vec{p}_H  d\vec{p}_{\bar{L}} d\vec{p}_{\bar{H}}
(2\pi)^4 \delta^{(4)}(p_L + p_H  - p_{\bar{L}} - p_{\bar{H}}) \times
\Bigg\{ \fHeq (1+\fHeq)  \\
& &  \left[ f_{L_\alpha}  \left(\sum_\beta (1-f_{\bar{L}_\beta})
|M_{sub}(L_\alpha H \to \bar{L}_\beta \bar{H})|^2  + 
\sum_{\beta\neq \alpha}(1-f_{{L}_\beta})
|M_{sub}(L_\alpha H \to {L}_\beta {H})|^2 \right) \right.
\nonumber \\
&-& \left. \left. (1-f_{L_\alpha}) 
\left(\sum_\beta f_{\bar{L}_\beta}
|M_{sub}( \bar{L}_\beta \bar{H} \to L_\alpha H)  |^2  + 
\sum_{\beta\neq \alpha} f_{{L}_\beta}
|M_{sub}(L_\beta H \to {L}_\alpha {H})|^2 \right) \right]
\right\} \nonumber ,
\eea
and 
\bea
\label{salb2}
S^{\bar \alpha}& = & \frac{1}{2E_L} 
\int  d\vec{p}_H  d\vec{p}_{\bar{L}} d\vec{p}_{\bar{H}}
(2\pi)^4 \delta^{(4)}(p_L + p_H  - p_{\bar{L}} - p_{\bar{H}}) \times
\Bigg\{ \fHeq (1+\fHeq)  \\
& & 
    \left[ f_{\bar L_\alpha} \left(\sum_{\beta \neq \alpha} (1-f_{\bar{L}_\beta})
|M_{sub}(\bar L_\alpha \bar H \to \bar{L}_\beta \bar{H})|^2  + 
   \sum_{\beta} (1-f_{{L}_\beta})
|M_{sub}(\bar L_\alpha \bar H \to {L}_\beta {H})|^2 \right) \right. 
\nonumber \\
&-& \left.\left.
(1-f_{\bar L_\alpha})
  \left(\sum_{\beta \neq \alpha} f_{\bar{L}_\beta}
|M_{sub}(\bar L_\beta \bar H \to \bar{L}_\alpha \bar{H})|^2  + 
   \sum_{\beta} f_{{L}_\beta}
|M_{sub}({L}_\beta {H} \to \bar L_\alpha \bar H)|^2 \right)\right] 
\right\} . 
\nonumber
\eea
Using the subtracted amplitudes as in the unflavoured case, the on-shell 
part of the subtracted amplitudes, \eq{osscat}, now is,
\bea
\label{osscatal}
S_{os}(L_\alpha H \rightarrow \bar{L}_\beta \bar{H}) =
 \frac{1}{2E_L} 
\int d\vec{p}_H d\vec{p}_{\bar{L}} d\vec{p}_{\bar{H}}
(2\pi)^4 \delta^{(4)}(p_L + p_H  - p_{\bar{L}} - p_{\bar{H}}) \qquad
\\
\times   f_{L_\alpha} \fHeq (1-f_{\bar{L}_\beta})(1+\fHeq) \ALHNal 
\frac{\pi \delta(s-M_{i}^2)}{M_{i}
\Gamma^{th}_{i}} \ANLHbbet \, . \nonumber
\eea
Continuing analogously to the unflavoured case, we obtain the Boltzmann
equations for the flavour asymmetries at first order in 
$\epsilon$ and $f_{\cal L}$:
\bea
\label{fullflavour}
\frac{\partial f_{\cal{L}_\alpha}}{\partial t}&-& {p}_L H 
\frac{\partial f_{\cal{L}_\alpha}}{\partial {p}_L} = \frac{1}{2E_L} 
\int d\vec{p}_H d\vec{p}_{N_i}
(2\pi)^4 \delta^{(4)}(\pN - p_L - p_H) 
\\ & \times& \left \{ \ADi \epial \, (f_{N_i} - \fNieq)
\left[ (1-\fLeq)(1+\fHeq) - \fLeq \fHeq \right]  - |A^i_{D,\alpha}|^2 \frac 1 2
f_{\cal{L}_\alpha}  (\fHeq+f_{N_i}) \right\}
\nonumber \\
&+& \frac{1}{2E_L} 
\int d\vec{p}_H d\vec{p}_{\bar L}d\vec{p}_{\bar H} (2\pi)^4 \delta^{(4)}( p_L
+ p_H - p_{\bar L} -p_{\bar H}) \fHeq ( 1 + \fHeq)\nonumber \\
&\times & \sum_{\beta} \left \{ 
    f_{\cal{L}_\beta} |M_{\alpha\beta}|^2_{off}  
  - f_{\cal{L}_\beta} |M_{\bar \alpha \beta}|^2_{off} 
  - f_{\cal{L}_\alpha} |M_{\alpha \beta}|^2_{off} 
  - f_{\cal{L}_\alpha} |M_{\bar \alpha \beta}|^2_{off} \right\} \ ,
\nonumber
\eea
where $|M_{\alpha \bar \beta}|^2_{off}$ is defined analogously to \eq{Moff}.
We have used CPT invariance, which 
implies that 
$|M(\bar L_\beta \bar H \to \bar{L}_\alpha \bar{H})|^2 =
|M(L_\alpha H \to {L}_\beta H)|^2$, etc., to write all the 
amplitudes with the flavour $\alpha$ as initial state, 
and also that the off-shell amplitudes are CP conserving 
at leading order in the couplings, so that 
$|M_{\alpha \bar \beta}|^2_{off} = |M_{\bar \alpha \beta}|^2_{off}$.
Finally, we have summed over all flavours $\beta$ 
in the last line of Eq.~(\ref{fullflavour}), since the extra 
contributions for $\beta = \alpha$
cancel between the first and the third terms.

Notice that in \eq{fullflavour}, we have kept also the off-shell
amplitudes, even taking into account that they are higher order in the Yukawas.
In fact, we expect $|M_{\alpha \bar \beta}|^2_{off}$ to be of order
$Y_{\alpha i}^2 Y_{\beta i}^2$ that is indeed smaller than
$Y_{\alpha i}^2$ unless $Y_{\beta i} \simeq 1$. However notice that some
of these substracted amplitudes contribute proportionally to
$f_{\cal{L}_\beta}$ and contributions $|Y_{\alpha i} Y_{\beta i}|^2
f_{{\cal L}_\beta}$  could be important with respect to the terms
$|Y_{\alpha i}|^2 f_{{\cal L}_\alpha}$ if 
$f_{{\cal L}\alpha}\ll f_{{\cal L}\beta}$ and/or 
$|Y_{\alpha i}| \ll |Y_{\beta i}|$.
Nevertheless, it is not clear a priori when these effects may play a 
relevant role.

\section{Full BEs in soft leptogenesis}
\label{appB}

Here, we describe in detail the derivation of 
the full BEs relevant for soft leptogenesis. 

Using $CPT$ invariance and  the definitions for the $CP$ 
asymmetries, the decay and inverse decay sneutrino amplitudes 
can be written as:
\be
\begin{array}{ccccc}
\Nis & = & \sbNi & \simeq &\displaystyle{\frac{1+\esi}{2}\As}  \; ,\\
\Nisb & = & \sNi & \simeq &\displaystyle{\frac{1-\esi}{2}\As} \; , \\
\Nif & = & \fbNi & \simeq &  \displaystyle{\frac{1+\efi}{2}\Af}  \; , \\
\Nifb & = & \ffNi & \simeq &\displaystyle{\frac{1-\efi}{2}\Af}  \; , \\
\end{array} 
\ee
where $\As$ ($\Af$) is the CP conserving tree-level sneutrino 
decay amplitude to scalars (fermions).

The evolution equation for the sneutrino distribution
with the same approximations as in Appendix A (i.e., 
to first order in $\efi,\esi,\mu_f,\mu_s$)
is given by
\bea
\bolsNi &=& \frac{1}{2E_{\widetilde{N}_i}} \int d\vec{p}_L  d\vec{p}_H 
(2\pi)^4
\delta^4({p}_{\widetilde{N}_i} -{p}_L -{p}_H) \nonumber \\
&\times & \left\{ \Af \left[ \fheq \fLeq (1+\fsNi) - \fsNi (1-\fLeq)(1-\fheq)
  \right] +
\right. \nonumber \\
& & + \left. \As \left[ \fHeq \fLteq (1+\fsNi) - \fsNi
    (1+\fLteq)(1+\fHeq) \right] \right\} \, . \label{BE_soft_snua}
\eea
For the lepton and slepton asymmetries we have:
\bea
\bollep & = & \sum_i
\left( D_i-\bar{D}_i \right)-2 S -
S_{L\widetilde{L}^{\dag}} + \bar{S}_{L\widetilde{L}^{\dag}} -
S_{L\widetilde{L}} + \bar{S}_{L\widetilde{L}} + S_g \, ,
\label{eq:bolf}\\
\bolslep & = & \sum_i\left( \widetilde{D}_i-\widetilde{D}_i^{\dag} 
\right)-2
\widetilde{S} - S_{L\widetilde{L}^{\dag}} +
\bar{S}_{L\widetilde{L}^{\dag}} + S_{L\widetilde{L}} -
\bar{S}_{L\widetilde{L}} + \widetilde{S}_g \, ,
\label{eq:bols}
\eea
where 
\bea
\label{Di}
D_i &=& \frac{1}{2E_L} 
\int d\vec{p}_{N_i} d\vec{p}_h  
(2\pi)^4 \delta^{(4)}(\pN - p_L - p_h)  \\
 & \times & \left\{ \fsNi (1-\fL)(1-\fheq) \Nif
- \fheq \fL (1+\fsNi)  \ffNi \right \}  \, , \nonumber 
\eea
\bea
\label{Dti}
\widetilde{D}_i &=& \frac{1}{2E_L} \int 
d\vec{p}_{N_i} d\vec{p}_{H} 
(2\pi)^4 \delta^{(4)}(\pN - p_{{L}} - p_{{H}})  \\
 & \times & \left\{ \fsNi (1+\fLt) (1+\fHeq) \Nis
- \fHeq \fLt (1+\fsNi)  \sNi \right \} \, , \nonumber 
\eea
and analogous expressions for $\bar{D}_i$ and 
$\widetilde{D}_i^{\dag}$, just changing particles by 
antiparticles.

As in the previous section, the scattering terms 
\bea
\label{sa}
S& = & \frac{1}{2E_L} 
\int  d\vec{p}_h  d\vec{p}_{\bar{L}} d\vec{p}_{\bar{h}}
(2\pi)^4 \delta^{(4)}(p_L + p_h  - p_{\bar{L}} - p_{\bar{h}}) \nonumber \\
& \times & 
      \left\{f_L \fheq (1-f_{\bar{L}})(1-\fheq)
|M_{sub}(L h \to \bar{L} \bar{h})|^2- \right. \nonumber \\ 
& - & \left. f_{\bar{L}} \fheq (1-f_L)(1-\fheq)
|M_{sub}(\bar{L} \bar{h} \to L h)|^2 \right\} \, ,   
\eea
\bea
\label{sta}
\widetilde{S} & = & \frac{1}{2E_L} 
\int  d\vec{p}_H  d\vec{p}_{\bar{L}} d\vec{p}_{\bar{H}}
(2\pi)^4 \delta^{(4)}(p_L + p_H  - p_{\bar{L}} - p_{\bar{H}}) \nonumber \\
& \times & 
      \left\{\fLt \fHeq (1+\fLtb)(1+\fHeq)
|M_{sub}(\widetilde{L} H \to \widetilde{L}^\dag H^\dagger)|^2
- \right. \nonumber \\ 
& - & \left. \fLtb \fHeq (1+\fLt)(1+\fHeq)
|M_{sub}(\widetilde{L}^\dagger {H}^\dagger \to \widetilde{L} H)|^2 
\right\} \, ,   
\eea
etc., are defined in terms of the subtracted amplitudes 
(see eqs.~(\ref{amplisub}),(\ref{amplios})). The thermal width of 
sneutrinos is 
$\Gamma^{i,th}= \Gamma^{i,th}_{f} + \Gamma^{i,th}_{s}$, 
with 
\be
\Gamma^{i,th}_{f} = \frac{1}{2 M_i} \int d\vec{p}_L d\vec{p}_H 
(2\pi)^4 \delta^{(4)}(p_{\widetilde{N}_i} - p_L - p_H) 
\left[(1-\fLeq)(1-\fheq) + \fLeq \fheq \right] \Af  \, ,
\ee
\be
\Gamma^{i,th}_{s} = \frac{1}{2 M_i} \int d\vec{p}_L d\vec{p}_H 
(2\pi)^4 \delta^{(4)}(p_{\widetilde{N}_i} - p_L - p_H) 
\left[(1+\fLteq)(1+\fHeq) + \fLteq \fHeq \right] \As  \, .
\ee

Finally, the scattering terms $S_g$, $\widetilde{S}_g$
correspond to the fast MSSM gaugino interactions 
$ L L \leftrightarrow \widetilde{L}  \widetilde{L} $
and are given by
\bea
\label{sg}
S_g& = & \frac{1}{2E_L} 
\int  d\vec{p}_1 d\vec{p}_{2} d\vec{p}_{3}
(2\pi)^4 \delta^{(4)}(p_L + p_1  - p_2 - p_3) \nonumber \\
& \times & 
      \left\{\fLt(p_2) \fLt(p_3) (1-f_L(p_L))(1-f_L(p_1))
|M (\widetilde{L}  \widetilde{L} \leftrightarrow L L )|^2- \right. 
\nonumber \\ 
& - & \left. f_L(p_L)) f_L(p_1) (1+\fLt(p_2)) (1+\fLt(p_3))
|M (L L \leftrightarrow \widetilde{L}  \widetilde{L})|^2 \right\} 
\nonumber \\
&- & \left\{{\rm particles \rightarrow antiparticles} \right\}
\, ,   
\eea
and analogously for $\widetilde{S}_g$.
Since these interactions are in equilibrium, we will not include 
them in our set of Boltzmann equations, but we shall impose that 
the chemical potentials for leptons and sleptons are equal,
$\mu_s=\mu_f$.

The derivation of the out-of-equilibrium condition is somehow lengthy, 
but completely analogous to the standard case described  
in appendix \ref{appA}, so we do not give many details here
(see for example \cite{ggr}).
The basic point is that 
at ${\cal O}(\eps)$, we can approximate the (s)lepton and
anti-(s)lepton distributions by the equilibrium ones, and   
then use the 
following relations between them
\bea
(1-\fheq)(1-\fLeq) & = & e^{E_{N_i}} \fsNieq \, 
\left[(1-\fheq)(1-\fLeq)+ \fheq \fLeq \right]\, , 
\\
(1+\fHeq)(1+\fLteq) & = & e^{E_{N_i}} \fsNieq \,
\left[(1+\fHeq)(1+\fLteq) + \fHeq \fLteq \right] \, ,
\eea
to reproduce the sneutrino thermal width, following the same 
procedure as in appendix \ref{appA}.
Finally, we obtain:
\bea 
\bollep & = & \sum_i
\frac{1}{2E_L} \int d\vec{p}_{\widetilde{N}_i} 
d\vec{p}_H (2\pi)^4 \delta(p_{\widetilde{N}_i} - p_L - p_H) \Af \nonumber \\
& \times & \left\{ \efi (\fsNi -\fsNieq)\left[ (1-\fLeq)(1-\fheq) +
    \fheq \fLeq \right] \right. - \label{BE_soft_Fa}
\nonumber \\
&  & \left. -  \frac 1 2 \flep (\fsNi + \fheq)\right\} 
\, , \\
\bolslep & = & \sum_i \frac{1}{2E_L} \int d\vec{p}_{\widetilde{N}_i}
d\vec{p}_H (2\pi)^4 \delta(p_{\widetilde{N}_i} - p_L - p_H)
\As \nonumber \\
& \times & \left\{ \esi (\fsNi -\fsNieq)\left[ (1+\fLteq)(1+\fHeq) +
    \fHeq \fLteq \right] +\right.
\nonumber \\
&  & + \left.  \frac 1 2 \fslep (\fsNi - \fHeq)\right\} \, , \label{BE_soft_Sa}
\\
\mu_f & = & \mu_s
\eea

\section{Analytic approximation in kinetic equilibrium}
\label{appC}
 
Our starting point are the Boltzmann equations for the (s)lepton
asymmetries (\ref{Ylep}) and (\ref{Yslep}), with the constraint $\mu_f
= \mu_s \equiv \mu $ imposed by fast gaugino interactions. 
In this approximation, we keep the phase space and statistical factors, 
crucial in 
soft leptogenesis, but we approximate the Fermi and Bose distributions 
by the Maxwell-Boltzmann one, since this is enough to obtain a 
non-vanishing CP asymmetry and allows to perform analitically 
the energy integrals.
After
integrating over the (s)lepton energy, we get
\bea
\frac{d\YLf}{dz} &=& 
 \frac{K_f z^2}{2\pi^2 s}  
\int_{z}^{\infty} d\e_N e^{-\e_N} 
 \left\{ - \epsilon \, \left( 
\frac{Y_{\widetilde{N}}}{Y_{\widetilde{N}}^{eq}}-1 \right)
\left[\lambda(1,x_L,x_h) y_N (1+2e^{-\e_N}) \right. \right.
\nonumber \\
& -& \left.\left.
 \left( e^{-\e_N(1-x_L+x_h)/2} +e^{-\e_N(1+x_L-x_h)/2} \right )
\left(e^{y_N \lambda(1,x_L,x_h) /2} - e^{-y_N \lambda(1,x_L,x_h) /2} \right) 
\right. \right ]
\nonumber \\
&- &  \left. \lambda(1,x_L,x_h) y_N
\mu \, T^2  \right \}
\\
\frac{d\YLs}{dz} &=& 
 \frac{K_s z^2}{2\pi^2 s}  
\int_{z}^{\infty} d\e_N e^{-\e_N} 
 \left\{  \epsilon \, \left( 
\frac{Y_{\widetilde{N}}}{Y_{\widetilde{N}}^{eq}}-1 \right)
\left[\lambda(1,x_{\sL},x_H) y_N (1+2e^{-\e_N}) \right. \right.
\nonumber \\
& +& \left.\left.
 \left( e^{-\e_N(1-x_{\sL}+x_H)/2} +e^{-\e_N(1+x_{\sL}-x_H)/2} \right )
\left(e^{y_N \lambda(1,x_{\sL},x_H) /2} - e^{-y_N \lambda(1,x_{\sL},x_H) /2} 
\right) \right. \right ]
\nonumber \\
&- &  \left. \lambda(1,x_{\sL},x_H) y_N
\mu \, T^2  \right \}
\eea
The washout term in the above equation is the standard one, 
for a non-vanishing (s)lepton mass, so, 
after integrating over $\e_N$ gives 
the Bessel function ${\cal K}_1(z)$. 
Notice that the chemical potential $\mu$ is related to the total
lepton asymmetry $Y_{{\cal{L}}_T} = Y_{\cal{L}} + Y_{\tilde{\cal{L}}}$
by
\be
\mu = \frac{\pi^2}{2} \frac{s}{m_L^2  {\cal K}_2(m_L/T) + m_{\sL}^2 {\cal K}_2(m_{\sL}/T)}
\YLT  \; .
\ee
Using that 
\be
e^{y_N \lambda /2} - e^{-y_N\lambda /2} 
= 2 \, \sinh \left(\frac{y_N \lambda}{2} \right) 
= 2 \sum_{n=0}^\infty  \frac{(y_N \lambda /2)^{2n+1}}{(2n+1)!} , 
\ee
we can perform the integral over $\e_N$, order by order in $y_N$. 
The resulting integrated Boltzmann equation for $Y_{{\cal{L}}_T} $
can be written as:
\be
\label{JN_mt}
\frac{d Y_{{\cal{L}}_T}}{dz} = 2 \, \epsilon \, K 
(Y_{\widetilde{N}}-Y_{\widetilde{N}}^{eq}) \frac{F_1(z)}{{\cal K}_2(z)} - 
\frac{K  z^3}{4} {\cal K}_1(z) F_2(z)Y_{\cal{L}_T} \, . 
\ee
with $F_1(z) = F_1^s(z)+F_1^f(z)$ and 
\bea
F_1^s(z) &=& \lambda(1,x_{\sL},x_H)
\left\{\lambda(1,x_{\sL},x_H) [{\cal K}_1(z) + {\cal K}_1(2z)] z
\right. \nonumber \\
& & \left. + f_1^s[z(3-x_{\sL}+x_H)/2] + f_1^s[z(3+x_{\sL}-x_H)/2] \right\}
\; ,\\
F_1^f(z) &=& (1-x_L+x_h) \lambda(1,x_L,x_h)
\left\{ - \lambda(1,x_L,x_h)  [{\cal K}_1(z) + {\cal K}_1(2z)]z
\right. \nonumber \\
& & \left. + f_1^f[z(3-x_L+x_h)/2] + f_1^f[z(3+x_L-x_h)/2] \right\} \; .
\eea
The function $f_1^s$ is given by 
\bea
f_1^s(z a) &=&   \sum_{n=0}^\infty  \frac{1}{2^{3n} n!}
\left(\frac{z}{a}\right)^{n+1}
[\lambda(1,x_{\sL},x_H)]^{2n+1}  {\cal K}_{n+1}(z a)
\nonumber \\
&\simeq& \frac {z}{a} \lambda(1,x_{\sL},x_H) {\cal K}_1(za) 
+ \frac 1 8 \left(\frac{z}{a}\right)^{2}  [\lambda(1,x_{\sL},x_H)]^3
{\cal K}_2(za) 
\nonumber \\
& & + \frac{1}{2^7}
\left(\frac{z}{a}\right)^{3}  [\lambda(1,x_{\sL},x_H)]^5
{\cal K}_3(za) + \ldots \; ,
\eea
and $f_1^f$ has the same structure, changing $\sL \rightarrow L$ and
$H \rightarrow h$.

When thermal masses are neglected, $F_1^s=F_1^f$, and the two 
different arguments of $f_1^{s,f}$ became the same ($3z/2$), 
leading to a global factor of 4 and the simplified function
$F^{(0)}(z)$ of \eq{F0}.
 
Finally, the function $F_2(z)$ is defined as
\be
\label{F2b}
F_2(z) =  2 T^2 \, \frac{ [\lambda(1,x_{L},x_h)]^2 (1-x_L-x_h) +
 [\lambda(1,x_{\sL},x_H)]^2}
{m_L^2  {\cal K}_2(m_L/T) + m_{\sL}^2 {\cal K}_2(m_{\sL}/T)}    \; ,
\ee

\end{document}